\documentclass[
  journal=largetwo,
  manuscript=article-type,
  year=2020,
  volume=37,
]{./style_files/cup-journal}

\usepackage{amsmath}                
\usepackage[nopatch]{microtype}     
\usepackage{booktabs}               
\usepackage{multirow}               
\usepackage{hyperref}               
\usepackage{xcolor}                 
\usepackage{url}                    
\usepackage{wrapfig}                
\usepackage{float}                  
\usepackage{./style_files/aas-macros}

\usepackage{colortbl}   

\title{Enhanced detection and identification of satellites using an all-sky multi-frequency survey with prototype SKA-Low stations}

\author{D. Grigg}
\affiliation{International Centre for Radio Astronomy Research, Curtin University, Bentley, WA, 6102, Australia}
\alsoaffiliation{DUG Technology, 76 Kings Park Rd, West Perth, 6005, WA, Australia}
\email[Grigg, D]{dylan.grigg@icrar.org}

\author{S. J. Tingay}
\affiliation{International Centre for Radio Astronomy Research, Curtin University, Bentley, WA, 6102, Australia}

\author{S. Prabu}
\affiliation{International Centre for Radio Astronomy Research, Curtin University, Bentley, WA, 6102, Australia}

\author{M. Sokolowski}
\affiliation{International Centre for Radio Astronomy Research, Curtin University, Bentley, WA, 6102, Australia}

\author{B. Indermuehle}
\affiliation{CSIRO Space \& Astronomy, PO Box 76, Epping, NSW, 1710, Australia}


\begin{document}

\begin{abstract}

With the low Earth orbit environment becoming increasingly populated with artificial satellites, rockets, and debris, it is important to understand the effects they have on radio astronomy. In this work, we undertake a multi-frequency, multi-epoch survey with two SKA-Low station prototypes located at the SKA-Low site, to identify and characterise radio frequency emission from orbiting objects and consider their impact on radio astronomy observations. We identified 152 unique satellites across multiple passes in low and medium Earth orbits from 1.6 million full-sky images across 13 selected ${\approx}1$ MHz frequency bands in the SKA-Low frequency range, acquired over almost 20 days of data collection. Our algorithms significantly reduce the rate of satellite misidentification, compared to previous work, validated through simulations to be $<1\%$. Notably, multiple satellites were detected transmitting unintended electromagnetic radiation, as well as several decommissioned satellites likely transmitting when the Sun illuminates their solar panels. We test alternative methods of processing data, which will be deployed for a larger, more systematic survey at SKA-Low frequencies in the near future. The current work establishes a baseline for monitoring satellite transmissions, which will be repeated in future years to assess their evolving impact on radio astronomy observations.

\end{abstract}

\section{Introduction}
Radio astronomy presents an opportunity to study the early Universe, as radio emission generated billions of years ago from neutral hydrogen is redshifted to low radio frequencies (100 - 200 MHz) \citep{eor_explanation}. These signals are expected to be weak, meaning that extremely sensitive instruments are needed to detect them and extract astrophysical information from vast data collections. In order to address this challenge, the most sensitive radio telescope of all time is being constructed, known as the Square Kilometre Array (SKA) \citep{ska}.

Civilisation also heavily utilises the radio spectrum for the purpose of television broadcast, FM radio, mobile phone communications, radio-navigation, and space-based communications, to name a few. In space-based communications, satellites transmit radio frequency energy to the surface of the Earth, to a ground station or device. If the satellite is transmitting information whilst passing through the receiving beam of a radio telescope, the strength of its transmission can be many orders of magnitude greater than the signals from even the strongest astronomical sources. This poses a significant and well-known risk for radio astronomy research \citep{rfi_gnss, radio_dynamic_zones, di_vruno, grigg_starlink}\footnote{\url{https://www.iau.org/static/science/scientific_bodies/working_groups/286/dark-quiet-skies-2-working-groups-reports.pdf}}.

The rate at which human-made objects are launched into space is also rising exponentially\footnote{https://ourworldindata.org/grapher/yearly-number-of-objects-launched-into-outer-space}. All of these need to be tracked to maintain a harmonious space environment, and the more objects there are, the more comprehensive the system tracking them needs to be. At the time of writing this paper, \texttt{space-track.org} tracks approximately 10,200 active satellites, 18,700 debris objects, and 16,700 objects which are variably tracked due to their small size or incomplete data, all in Earth orbit. The intrinsic risk of interference at radio frequencies from satellites operating in Earth orbit is increasing, simply by virtue of the greatly increased numbers of operating objects.

The International Telecommunication Union (ITU) manages satellite spectrum allocations in the radio band under its radiocommunication sector (ITU-R)\footnote{\url{https://www.itu.int/en/ITU-R/information/Pages/default.aspx}}. In terms of protections for radio astronomy from Earth, only approximately 3.7\% of the spectrum allocations in the frequency range 50 - 350 MHz receive some protection for radio astronomy. Primary or secondary allocations for radio astronomy exist at 73.00 - 74.60 MHz, 150.05 - 153.00 MHz, and 322.00 - 328.60 MHz across specific geographic regions in the ITU-R radio regulations\footnote{\url{https://www.itu.int/pub/R-HDB-22-2013}}. Footnote 5.149 in this document also states that ``administrations are urged to take all practicable steps to protect the radio astronomy service from harmful interference'', and applies to the frequency ranges 73.00 - 74.60 MHz and 322.00 - 328.60 MHz.

In the 50 - 350 MHz band, several allocations to the mobile-satellite, meteorological-satellite, and amateur-satellite services exist in the radio regulations, both as outright allocations and under footnote 5.254$^{\text{d}}$ for defence use. These can significantly impact telescopes such as the SKA, for which the low-frequency half (SKA-Low) is being built in Western Australia.  This is due to the historical nature of the regulations and the historical claims to the radio astronomy spectrum being limited, and also due to the rapid acceleration of satellite and launch technologies.

Historically, radio telescopes have successfully managed the limited regulatory protection of the spectrum for radio astronomy by operating from remote areas, largely free from Radio Frequency Interference (RFI). Examples include the Murchison Widefield Array \citep[MWA,][]{mwa} and Australia Square Kilometre Array Pathfinder \citep[ASKAP,][]{askap} both in Western Australia, the MeerKAT telescope \citep{meerkat} in South Africa, and the Very Large Array \citep{vla} in New Mexico, USA. With satellites essentially visible from anywhere on the Earth, geographic isolation does not mitigate against RFI originating from satellites, highlighted by work using the MWA and Engineering Development Array 2 \citep[EDA2,][]{mwa, stevep_2, tingay_eda_ssa, grigg}.

\begin{figure}[hbt!]
\centering
\includegraphics[width=1\linewidth]{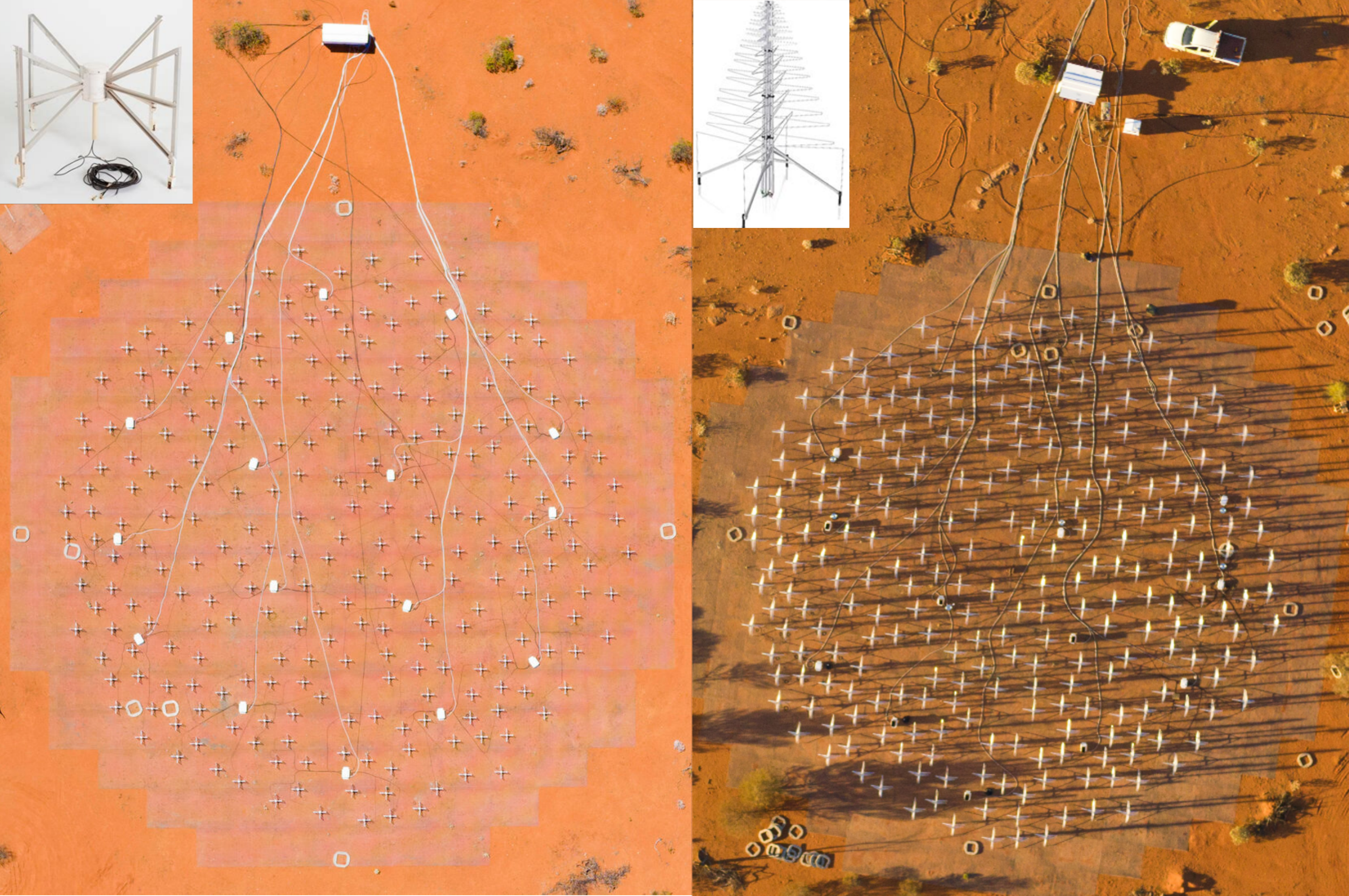}
\caption{A comparison of the EDA2 (left) and AAVS2 (right) low-frequency radio telescopes used in this work.}
\label{fig:eda2_aavs2}
\end{figure}

The SKA-Low telescope has a planned observing frequency of 50 - 350 MHz, and satellites may impact the pursuit of its headline science goals. This type of scenario is already a reality. An example is the ALMA radio telescope in Chile, which needed to introduce restrictions on observing at the same frequency as CloudSat transmissions, as there was a non-negligible risk of damage to its receivers\footnote{\url{https://legacy.nrao.edu/alma/memos/html-memos/alma504/memo504.pdf}}.

The work we describe in this paper is a first step toward characterising the impact of all satellites, specifically across the full SKA-Low frequency range of 50 - 350 MHz at the SKA-Low site. In this work, we test various methods of acquiring and processing data to understand the optimal way to conduct a more comprehensive survey in the future.

The radio telescopes we use are the Aperture Array Verification System, version 2 \citep[AAVS2,][]{aavs2} and the Engineering Development Array, version 2 \citep[EDA2,][]{eda2}. These are pictured in Figure \ref{fig:eda2_aavs2}. The two SKA-Low prototype stations were chosen because they are representative of a single SKA-Low station. The two telescopes are on the same site as the SKA-Low in Western Australia, Inyarrimanha Ilgardi Bundara, the CSIRO Murchison Radio-astronomy Observatory \footnote{\url{https://research.csiro.au/mro/}}. This site has been designated a Radio Quiet Zone by national legislation \citep{rqz}. This makes the AAVS2 and EDA2 ideal for understanding the environment in which the SKA-Low will operate, especially as it pertains to sources of RFI.

Two methods are explored to detect satellites with a radio telescope. The first is the detection of direct transmissions from the satellite. The mechanisms behind these types of radio emissions are defined as follows: `intentional electromagnetic radiation' (IEMR) is intentional transmission from a satellite's antenna at its designated downlink frequency; `interference' is any additional energy from the satellite's transmission antenna that is outside the designated downlink frequency; and `unintentional electromagnetic radiation' (UEMR) is energy emitted from another component of the satellite that is not the antenna (for example the propulsion system). IEMR is regulated by the ITU's Radio Regulations (ITU-R)\footnote{\url{https://www.itu.int/pub/R-HDB-22-2013}}, whereas interference and UEMR are not currently regulated by any international regulatory bodies. As per the ITU-R definition, anything which is IEMR is henceforth referred to as `emission'.

\begin{figure*}[hbtp]
\centering
\includegraphics[width=1\linewidth]{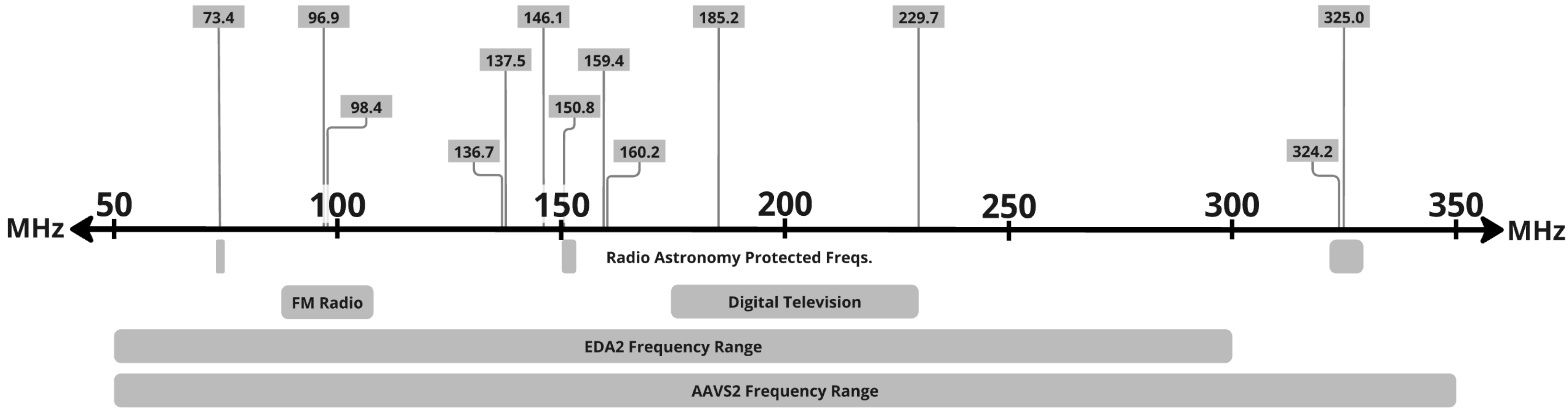}
\caption{The 15 frequencies of datasets used in this work alongside noteworthy spectrum allocations.}
\label{fig:freq_allocation}
\end{figure*}

The second method is the detection of satellites via the reflection of terrestrial sources of radio emission. This is a form of radar, and in radar parlance is described as passive non-coherent bi-static radar \citep{passive_radar}. It is passive, meaning that no signal is deliberately emitted by the observer to make a detection, and is bi-static, meaning that the source (illuminator) and the receiver are spatially separated. It is non-coherent because the received signal is not compared to reference signals (the illumination signal) to achieve phase coherence in order to allow for a range measurement.  

One prominent source of illumination in this scenario is commercial or private FM radio transmitters distributed throughout Australia, which can output up to 250 kW of power over 0.2 MHz of bandwidth\footnote{\url{https://www.acma.gov.au/list-transmitters-licence-broadcast}}. This transmission is directed over the land, but due to complex transmission beams, significant energy can be directed away from the Earth \citep{hennessy}. The radio waves can reflect off objects in the local airspace or in Earth orbit and subsequently be detected by radio telescopes operating in this frequency range. Other sources of illumination are also possible, for example DTV broadcasting, and many other licenced sources of terrestrial and airborne transmission.

Previous studies have developed processes for the detection of satellites in AAVS2 and EDA2 datasets, based on both these methods of detection. An initial study by \citet{tingay_eda_ssa} began by acquiring three days worth of ${\approx}$1 MHz bandwidth data overlapping several known FM transmitters using the EDA2 and making seven ``by eye'' detections. A follow up study by \citet{grigg} developed and utilised autonomous detection algorithms on the same data set, resulting in a 6 fold increase in the detection rate. Work by \citet{soko_eda2} used the AAVS2 and EDA2 in tandem over a small selection of frequencies within the SKA-Low observing bandwidth and detected 30 unique objects (satellites or debris) as part of a larger survey of RFI at the site. Most recently, \citet{grigg_starlink} detected IEMR and UEMR from Starlink satellites using the EDA2, from some of the same datasets that will be reported on in the current paper. Separately to these, \citet{di_vruno} and \citet{starlink_bassa} also detected UEMR from Starlink satellites using the LOFAR radio telescope \citep{lofar_desc}.

Building on our previous work using AAVS2 and EDA2, our current work uses twelve frequencies (${\approx}$1 MHz bandwidth per frequency) which are listed in Table \ref{tab:survey_table} and illustrated in Figure \ref{fig:freq_allocation}. These frequencies were selected based on a range of factors, in an attempt to span the full SKA-Low frequency range at discrete frequency intervals. Our purpose is to take the first step in systematically surveying the full SKA-Low frequency range, in order to understand the impacts that RFI from satellites (and other airborne objects such as aircraft) may have on the SKA-Low. 

Improving autonomy of the detection algorithms to delineate between satellites and RFI is one of the key goals of this work, which is quantified with a simulation to estimate a misidentification rate. We explore data processing and signal identification methods, and compare them with other recent studies \citep{grigg, soko_eda2}. This study will be a key step in informing the design of a larger, more comprehensive survey in the future. This survey will also be a baseline against which to compare future monitoring of RFI from satellites.

Section \ref{methods} presents the methods we have used to collect and process data from the AAVS2 and the EDA2. Section \ref{results} presents our results, and discussion of the results is in Section \ref{discussion}. In Section \ref{conclusion} we draw our conclusions from this work.

\section{Methods}
\label{methods}

This study was designed to utilise frequency channels within the SKA-Low frequency range of 50 - 350 MHz. Frequency channels which were known to be contaminated with RFI (from satellites, aircraft, meteors, and terrestrial transmitters) were intentionally chosen, along with others which were thought to contain little RFI. This section will examine the choice of frequencies, data processing techniques, and a simulation used to quantify the expected misidentification rate.

\subsection{Instrumentation}

The two radio telescopes used for this work were the AAVS2 \citep{aavs2} and EDA2 \citep{eda2}. They both feature an arrangement of 256 antennas within a diameter of ${\approx}$38 m and ${\approx}$35 m, respectively. The difference between the two telescopes is that the AAVS2 uses SKA prototype antennas (known as SKALA) and the EDA2 uses MWA bowtie antennas. The functional bandwidth of the two telescopes is comparable to the SKA-Low being approximately 50 - 350 MHz.

In previous internal tests, it was found that the AAVS2 had a higher sensitivity at the horizon. This increased the detrimental effect of transmission directly from terrestrial radio transmitters, resulting in a higher percentage of unusable data. The AAVS2 was therefore only used to acquire higher frequency data, making use of its wider frequency range compared to the EDA2.

\subsection{Choice of frequencies}

The choice of specific frequencies for this work was driven by their practical significance and scientific relevance. These are listed in Table \ref{tab:survey_table}, and illustrated in Figure \ref{fig:freq_allocation}. In total, almost 20 days of telescope recording time was collected, comprising 1.6 million images over a cumulative total of ${\approx}$12 MHz of bandwidth.

The `notes' column in Table \ref{tab:survey_table} describes the reasoning behind the choice of each frequency. These will be briefly explained. 

\begin{table*}
\centering
\caption{Surveyed frequencies}
\label{tab:survey_table}
\begin{tabular}{|c|c|c|c|c|c|c|c|}
\hline
\begin{tabular}[c]{@{}c@{}}\textbf{Central} \\ \textbf{frequency (MHz)}\end{tabular}
& \begin{tabular}[c]{@{}c@{}}\textbf{Start date} \\ \textbf{(UTC)}\end{tabular} 
& \textbf{Duration} 
& \begin{tabular}[c]{@{}c@{}}\textbf{Integration} \\ \textbf{time (s)}\end{tabular} 
& \textbf{Telescope} 
& \begin{tabular}[c]{@{}c@{}}\textbf{Channel} \\ \textbf{bandwidth (MHz)}\end{tabular} 
& \begin{tabular}[c]{@{}c@{}}\textbf{Number of} \\ \textbf{channels}\end{tabular} 
& \textbf{Notes} \\
\hline
73.4  & 2023-05-28 20:47:20.5 & 37 min       & 1   & EDA2  & 0.0289 & 32  & ITU-R protected\\ 
96.9  & 2022-12-16 10:10:30.5 & 6 min        & 1   & EDA2  & 0.0289 & 32  & ISS pass\\
96.9  & 2022-12-18 18:22:15.5 & 7 min        & 1   & EDA2  & 0.0289 & 32  & ISS pass\\
98.4  & 2022-10-18 18:25:12.5 & 5 min        & 1   & EDA2  & 0.0289 & 32  & ISS pass\\
136.7 & 2023-04-25 01:04:37.5 & 12 min       & 1   & EDA2  & 0.0289 & 32  & SOLRAD 7B pass\\
137.5 & 2023-03-17 12:19:09.0 & 19 h 37 min  & 2   & EDA2  & 0.926  & 1   & Known downlink freq.\\
146.1 & 2021-11-18 01:08:33.0 & 24 h 22 min  & 2   & EDA2  & 0.926  & 1   & Archive\\
150.8 & 2023-05-30 20:05:21.5 & 4 min        & 1   & EDA2  & 0.0289 & 32  & ITU-R protected\\
150.8 & 2023-06-13 12:56:59.5 & 2 h 00 min   & 1   & EDA2  & 0.0289 & 32  & ITU-R protected\\
159.4 & 2020-06-26 12:34:07.0 & 132 h 35 min & 2   & AAVS2 & 0.926  & 1   & \citet{soko_eda2}\\
159.4 & 2020-06-26 11:48:10.0 & 133 h 38 min & 2   & EDA2  & 0.926  & 1   & \citet{soko_eda2}\\
159.4 & 2021-11-16 01:24:18.0 & 22 h 57 min  & 2   & EDA2  & 0.926  & 1   & Monitoring\\
160.2 & 2023-06-23 05:40:15.0 & 55 h 04 min  & 2   & EDA2  & 0.926  & 1   & Archive\\
185.2 & 2020-02-07 10:44:49.3 & 39 h 59 min  & 4.5 & EDA2  & 0.926  & 1   & Digital television\\
229.7 & 2021-11-16 01:10:45.0 & 23 h 14 min  & 2   & AAVS2 & 0.926  & 1   & Monitoring\\
324.2 & 2023-05-28 20:02:08.0 & 4 h 11 min   & 2   & AAVS2 & 0.926  & 1   & ITU-R protected\\
324.2 & 2023-06-13 11:47:02.0 & 2 h 37 min   & 2   & AAVS2 & 0.926  & 1   & ITU-R protected\\
325.0 & 2023-05-30 19:20:09.0 & 7 h 59 min   & 2   & AAVS2 & 0.926  & 1   & ITU-R protected\\
\hline
\end{tabular}
\end{table*}

We choose frequencies at 73.4 MHz, 150.8 MHz, 324.2 MHz, and 325 MHz to overlap with ITU-R protected frequencies for radio astronomy to search for satellite transmissions or UEMR.

Work by \citet{soko_eda2} had previously identified satellites at frequencies of 159.4 MHz and 229.7 MHz. These were again chosen to compare satellite identifications in our study with Sokolowski et al.'s work. Differences in the number and type of detections made in these two studies prompted a reprocessing of two ${\approx}$130 hour datasets acquired by the AAVS2 and EDA2 from Sokolowski et al.'s paper (on 2020-06-26 both at 159.4 MHz) to determine if our new algorithms result in a higher detection rate and lower misidentification rate. Note that a STARLINK train has been discovered transmitting UEMR in the 2021-11-16 159.4 MHz dataset, but is omitted from this analysis as it has already been reported by \citet{grigg_starlink}.

It is interesting to understand whether radio astronomy research can be performed at or near frequencies which are allocated to satellite transmission. The frequency of 137.5 MHz was chosen for this reason, as many satellites have their designated downlink frequency within this band. This same dataset was examined in the recent paper by \citet{grigg_starlink}, only describing the STARLINK satellite transmission. We now describe the other non-STARLINK detections in this dataset.

It is known that FM radio waves reflect off satellites \citep{steve, hennessy, tingay_eda_ssa}, but it is unknown whether reflections of digital television transmission can also be detected. The frequency of 185.2 MHz was chosen to overlap with four DTV transmitters in Western Australia, which are listed in Table \ref{tab:transmitters}. Note that satellite detections in the FM band had been described in \citet{grigg}, which was why longer observational periods were not acquired at FM frequencies.

The two frequencies 146.1 MHz and 160.2 MHz were pre-existing in the EDA2 data archive and were made available for this work.

Three frequencies were also chosen to test a new method of image processing called `frequency differencing', explained in more detail in Section \ref{freq_dif}. The 96.9 MHz and 98.4 MHz datasets were chosen to coincide with passes of the International Space Station (ISS; NORAD 25544) at frequencies which overlapped the terrestrial FM transmitters listed in Table \ref{tab:survey_table}. The 136.7 MHz dataset coincides with a pass of the satellite SOLRAD 7B (NORAD 1291). Instead of recording data at a single `coarse' 0.926 MHz frequency channel, these four datasets were captured using 32 `fine' frequency channels (28.9 kHz bandwidth per channel). As the 32 fine channels could be averaged into a single 0.926 MHz coarse channel, these datasets were also used in the analysis in Section \ref{data_processing}.

\begin{table}
\centering
\caption{Notable terrestrial transmitters. Nominal bandwidth of the transmitters is 0.2 MHz.}
\label{tab:transmitters}
\begin{tabular}{|c|c|c|c|}
\hline
\textbf{Location} 
& \begin{tabular}[c]{@{}c@{}}\textbf{Distance from} \\ \textbf{telescope (km)}\end{tabular} 
& \textbf{Frequency}
& \textbf{Power (kW)} \\
\hline
\multicolumn{4}{c}{FM transmitters $\geq$10 kW} \\
\hline
Northam               & 553     & 96.5    & 10   \\
Geraldton             & 295     & 96.5    & 30   \\
Southern agricultural & 889     & 96.9    & 80   \\
Perth                 & 593     & 96.9    & 100  \\
\hline
Geraldton             & 295     & 98.1    & 30   \\
Central agricultural  & 589     & 98.1    & 80   \\
Perth                 & 593     & 98.5    & 16   \\
\hline
\multicolumn{4}{c}{Digital television transmitters $\geq$1 kW} \\
\hline
Morawa                & 301     & 184.5   & 3.2  \\
Kalgoorlie            & 644     & 184.625 & 4    \\
Perth 1               & 593     & 184.5   & 50   \\
Perth 2               & 593     & 184.5   & 50   \\
\hline
\end{tabular}
\end{table}

\subsection{Data Processing}
\label{data_processing}

Section \ref{data_processing} is focused on the analysis using the time differencing technique, which is described below.

\subsubsection{Pre-processing}
\label{preprocessing}

The antenna voltages acquired from the AAVS2 and EDA2 were cross-correlated for all baselines. Table \ref{tab:survey_table} lists the correlated integration time for each observation, with each acquisition capturing 0.926 MHz of total bandwidth. For the single channel datasets, these visibilities output by the correlator were calibrated using the procedure described by \citet{soko_eda2}, which is an extension of the procedure used in \citet{eda2_old_calibration}. This calibration corrects for phase variations (due to residual unaccounted for cable delays), and amplitude (to bring to an absolute flux density scale). It also assumes a quiet Sun model, and while this may not give the highest level of accuracy near solar maximum, it is adequate for this analysis where the astrometry of the signals is the most important factor.

Subsequent to calibration, visibilities were then imaged and cleaned with the radio astronomy data processing package \texttt{MIRIAD} \citep{miriad}, for which the procedure is outlined in the work of \citet{tingay_eda_ssa}. This produced full-sky images in both the XX (east-west oriented) and YY (north-south oriented) orthogonal polarisations.

For the four datasets with 32 channels, a different correlation package was used to handle the fine channelisation. This meant that the only difference with the processing to the image stage was that amplitude calibration was not performed. A simplified amplitude calibration was performed instead at the image stage to allow the amplitudes in these datasets to be comparable to the single channel datasets. This will be briefly explained.

The 2022-12-18 96.9 MHz dataset had Hydra A visible in the images. The Hydra A beam corrected flux density was fitted for in each polarisation, and using the polynomial correction function and coefficients measured by \citet{calib}, the measured image intensities were multiplied by the correction factors for each polarisation. The 2022-12-16 96.9 MHz dataset had no astronomical radio sources visible in the dataset, which meant that the same correction had to be applied from the dataset two days earlier. The 2022-10-18 98.4 MHz dataset had Fornax A visible, and was also corrected for with the same method, also using coefficients measured by \citet{calib}. The last dataset at 136.7 MHz had the Sun visible in the images. The same quiet Sun model from \citet{tingay_eda_ssa} was used to derive the bulk correction applied to the images for each polarisation.

\begin{figure}
\includegraphics[width=1\linewidth]{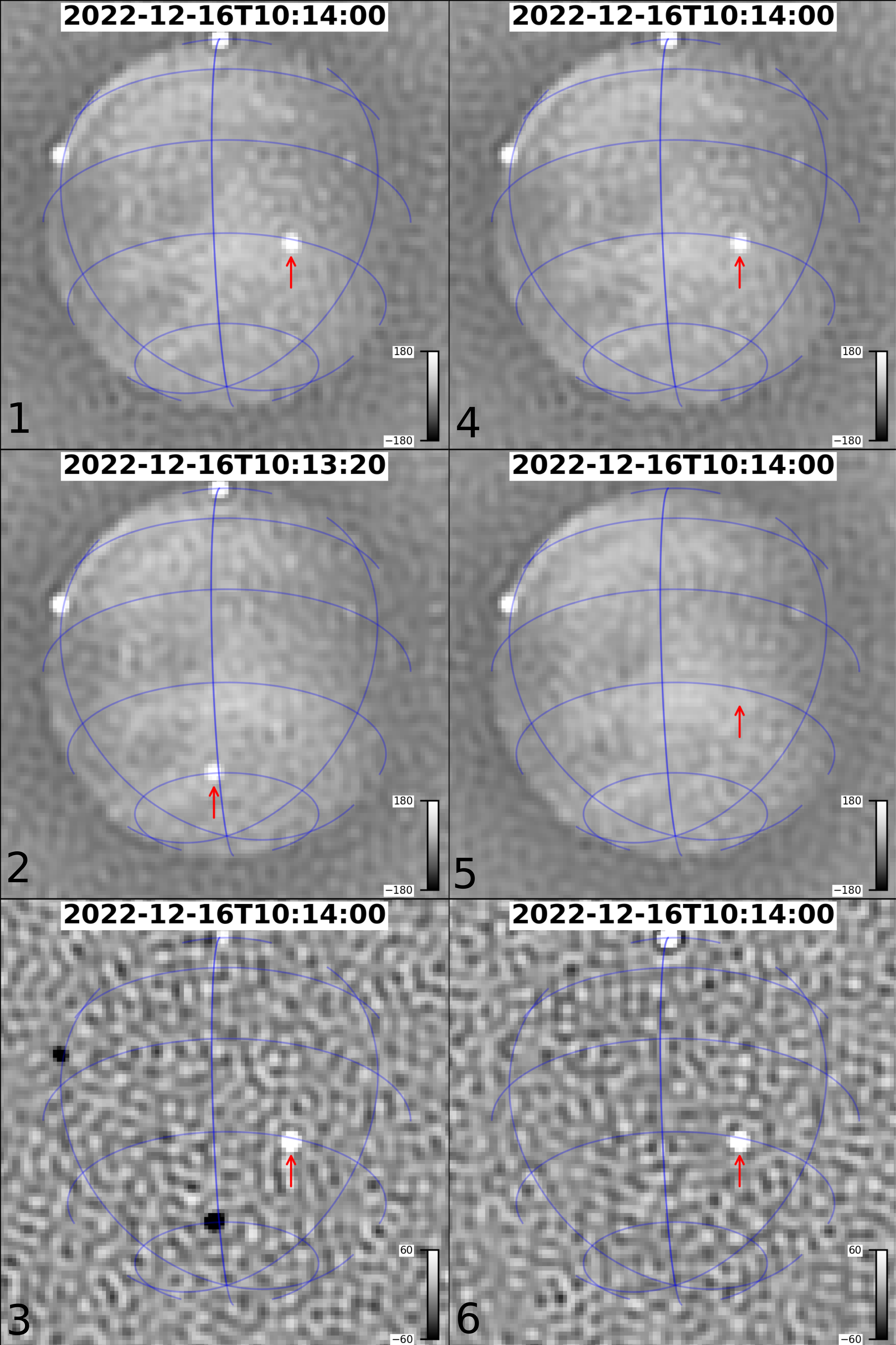}
\caption{The time and frequency differencing techniques are demonstrated with these images of the full sky, encompassed by the blue grid. The red arrows show the location of the ISS. For time differencing, image two is subtracted from image one with 40 s separating the two images with a central frequency of 96.9 MHz over a 28.9 kHz bandwidth. This gives image three which is the `time differenced' image, and clearly shows the separation of the reflected FM transmission from the ISS. For frequency differencing, image four (96.90 MHz) shows the ISS, while in image five (97.09 MHz) the ISS is not visible (both have a 28.9 kHz bandwidth). Image five is subtracted from image four to give the `frequency differenced' image shown in image six. It is important to note that the time difference method retains the negative subtracted ISS from the previous frame.}
\label{fig:dif_images}
\end{figure}

For the primary satellite detection and identification, a technique called `time differencing' was used to process the images. Satellites move very quickly over the sky compared to the movement of astrophysical sources, from the point of view of an observer on the Earth. Low Earth orbit (LEO) satellites can for example can pass overhead up to $\approx$1 degree/second, while the sky moves up to 15 degrees/hour. Therefore images of the sky, taken close to each other in time, are differenced to isolate the signals which traverse the sky quickly. This is shown in the left column of Figure \ref{fig:dif_images} and is a technique which has been used in previous similar studies \citep{tingay_eda_ssa, grigg, grigg_starlink}. Other variations in amplitudes between the two images will also be visible in the time differenced images. Examples include other transmitters which vary in directionally transmitted power, and scintillation of astrophysical sources.

All datasets in Table \ref{tab:survey_table} were time differenced with a nominal separation of 40 s between the two differenced images, for every image other than the first 40 s of the dataset. This allowed a satellite in low earth orbit to be sufficiently separated in the two images used for differencing. Image three in Figure \ref{fig:dif_images} shows this effect with the positive (white) amplitude emission being separated from the subtracted negative (black) amplitude emission. This met a trade off between the satellite's emission being separated enough to not overlap, whilst not introducing artefacts from imperfect background subtraction due to the movement of the sky between the two images. The satellite detection and identification algorithms were performed on these time differenced images.

In this paper, we refer to `candidates' as the detection of some signal for which the origin is not yet certain. This could be from a satellite, or noise, or any high flux density signal which is detected by the algorithms outlined in Section \ref{making_detections}. The candidates are then subject to a sorting algorithm which takes all of the candidates and algorithmically sorts them to determine if any are likely to be caused by a satellite. If a specific satellite is attributed to a given detection then we call it an identification. The identification algorithms are described in Section \ref{making_identifications}.

\subsubsection{Candidate Detections}
\label{making_detections}

A brute force fitting algorithm was applied to the images to make detections. The XX and YY orthogonal polarisations were run separately through this algorithm. The XX polarisation images are more sensitive to the North and the South, and the YY images are more sensitive to the East and West. If these were combined to form a Stokes I image, that extra sensitivity to the horizon due to one polarisation would be averaged out by the lack of sensitivity in the other, which is the reason the two polarisations are kept separated.

The satellite positional information used in this work was in the form of two line element sets (TLEs)\footnote{Sourced from \texttt{space-track.org}}, which describe six instantaneous Keplerian orbital elements at an epoch. Using the SGP4 propagation routine in the Skyfield python library \citep{skyfield}, these TLEs were used to calculate the positions of satellites for a specific image. All TLEs between two days either side of the start and end times of a given survey were collected. A perigee limit of 2,000 km was placed on the TLEs, as satellites in higher orbits would be moving too slow to be visible in the time range of the difference images.

The brute force fitting then proceeds as follows: for each time differenced image, the pixel locations of all satellites visible above the horizon were calculated. An elevation cutoff of 20$^\circ$ was imposed to prevent transmitters on the horizon being detected, as well as limiting the effect of photometric centroid shift which is explained further in \ref{app:sinusoidal}. The two dimensional elliptical Gaussian function shown in Equation \ref{eq:gaussian} is a reasonable approximation of the primary lobe of the point spread function (PSF) of the AAVS2 and EDA2. An elliptical Gaussian function is needed, as a LEO satellite at a higher frequency can move further than the width of the PSF.

For example, a satellite at a zenithal range of 400 km will travel 2.2$^\circ$ in 2 s (the integration time of the image). At a frequency of 328.1 MHz, the width of the PSF is $\approx1.5^\circ$. This motion, convolved with the PSF will create a streak, which when fitted with an elliptical Gaussian has fitting uncertainties two orders of magnitude below the 5$\%$ threshold.

This Gaussian does not need to be normalised, as the peak flux density of the fitted Gaussian gives the flux density of the received signal. An attempted fit for this function at the location of every satellite on the sky is made, meaning that all sources of radio emission at that frequency will be fit, regardless of their origins.

\begin{equation} \label{eq:gaussian}
I_{\text{model}} = A \cdot \exp\left(-\left[\frac{(x - x_{0})^{2}}{2 {\sigma_{x}}^2} + \frac{(y - y_{0})^{2}}{2 {\sigma_{y}}^2}\right]\right) ,
\end{equation}

The fitting algorithm used was from the \texttt{Scipy} library \citep{scipy}, which in this case estimates the Gaussian's amplitude ($A$), x coordinate ($x_{0}$), y coordinate ($y_{0}$), and the standard deviation of the Gaussian in the y direction ($\sigma_{y}$), which is related to the full width at half maximum (FWHM) of the Gaussian by FWHM $= 2\sigma\sqrt{2ln(2)}$. $\sigma_{x}$ is known from the width of PSF at the given frequency, and as such is not fitted for in this equation. An estimate of each of these was provided to constrain the fits, with $A$ estimated to be the maximum pixel value in a 3x3 kernel centred on $x_{0}$ and $y_{0}$, the predicted TLE pixel values of the satellite, and $\sigma_{y}$ was estimated from the width of the PSF. The standard deviation on each fitted parameter was calculated by taking the square root of the diagonalised covariance matrix of the fit. These standard deviations were treated as a measure of the uncertainty on each of the fitted parameters.




If a fit was successful for a given TLE position of a satellite, it had to pass through several constraints before it was deemed a candidate:
\begin{enumerate}
    \item $A >$ 1 Jy/beam;
    \item $A_{err} > 0.05 A$ and $\sigma_{err} > 0.05 \sigma$;
    \item The separation between the TLE predicted position and the fitted position could not be greater than the larger of two pixels or three degrees. Three degrees can be a fraction of a pixel at lower elevations, which is why a pixel and degree threshold is given; and
    \item If the RMS of the difference image was less than 10,000 Jy/beam (meaning that radio galaxies and the Sun were likely visible if overhead), the coordinates of the fit could not be inside an exclusion radius around the Sun and other radio galaxies which was defined as a function of the point spread function width at that frequency.
\end{enumerate}
The final part of this first processing step was to perform a beam correction on the detected flux density of the candidate for each polarisation. This was calculated by dividing $A$ by the value of the normalised antenna beam at the azimuth/elevation location of the candidate for each polarisation.

\subsubsection{Identifications}
\label{making_identifications}

In this Section, a qualitative description of how identifications are made from the list of candidates is described, with the reproducible mathematical steps and quantitative analysis available in \ref{app:identifications}.

For each survey, the list of candidates needed to be sorted to ensure satellites could be correctly identified. In this list, there may be multiple candidates detected at the same location, at the same time. This needs to be reduced to one candidate (for an identification) or zero candidates (for RFI). We now refer to the TLE predicted position of a satellite as the `predicted' position, and the position of the fitted Gaussian ($x_{0}$ and $y_{0}$) as the `measured' position. Each candidate will have this predicted and measured position information.

This proves to be valuable, as when the direction of travel of the predicted and measured positions is similar over time, it increases the probability of a correct identification. If the predicted and detected positions are travelling in different directions over the sky, then these candidates can be discarded. Therefore, knowing the instantaneous direction of travel of multiple candidates potentially associated with the same satellite can be used to discard infeasible candidates for that satellite.

The list of candidates is then sorted so all candidates for each satellite are put into 15 minute bins which are designed to be long enough that any object in the TLE list will pass over the sky completely during this time. The TLE list includes satellites which have a perigee up to 2,000 km. A satellite at a range of 2,000 km at the zenith would take at a maximum nine minutes to travel from 20$^\circ$ elevation (the elevation cutoff) on one side of the sky to 20$^\circ$ elevation on the other side of the sky. The 15 minute window ensures overlapping bins are not necessary. The total number of candidates in each bin ($N$) is stored with the individual candidates. If there are candidates for more than one satellite at the same location at the same time, the candidate with the most number of detections across the pass is kept, with the others being discarded.

\begin{figure}[hbt!]
\centering
\includegraphics[width=1\linewidth]{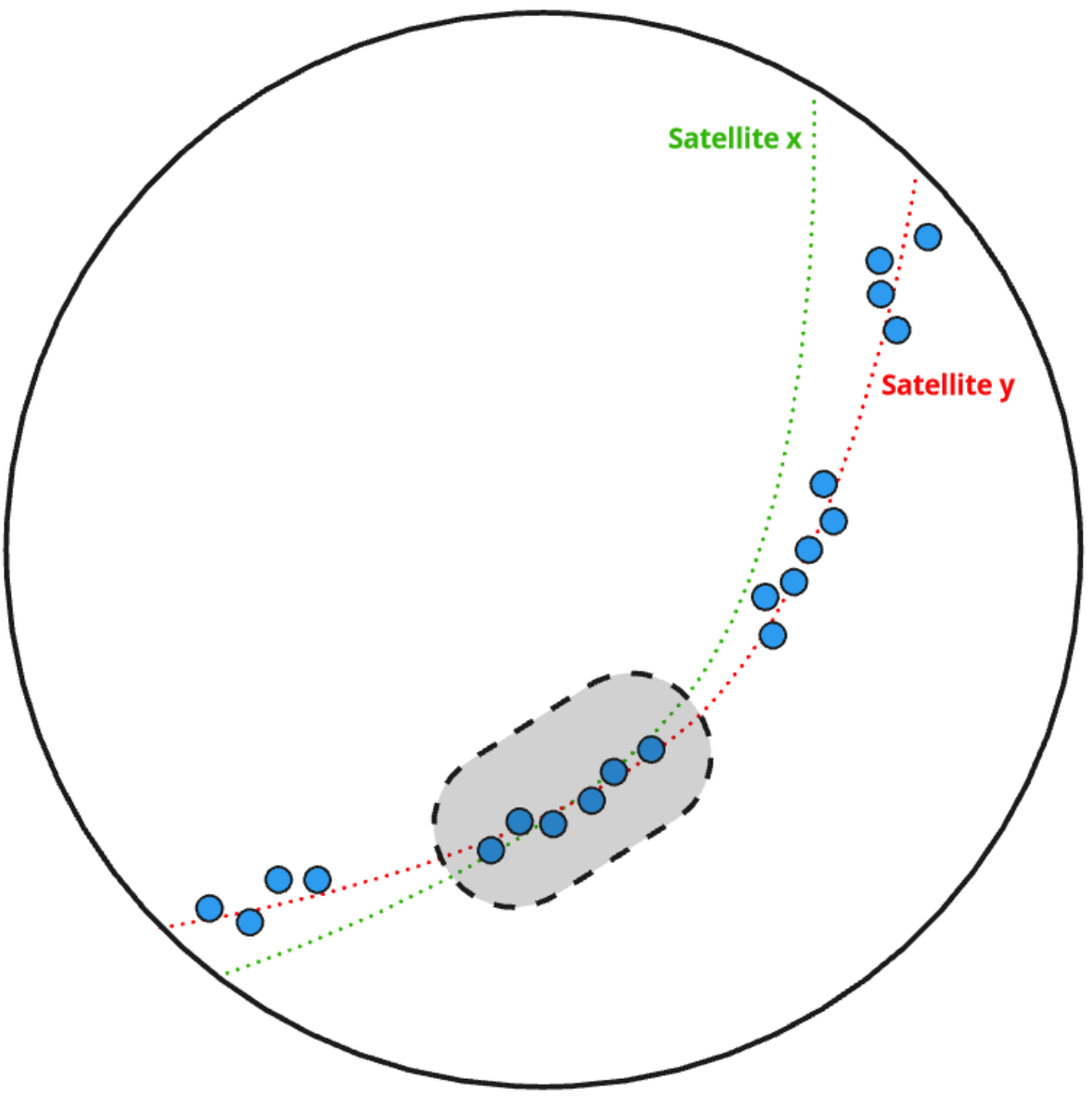}
\caption{An illustration of a scenario where the predicted paths of two satellites `x' (green) and `y' (red) are similar over the sky. The blue points are the measured positions of satellite y in 20 separate images. The six measured positions in the shaded region are flagged as candidates for both satellites.}
\label{fig:sat_diagram_path_comparison}
\end{figure}

Figure \ref{fig:sat_diagram_path_comparison} shows an example of this. In this figure, satellite `y' transmits twenty times as it passes over the visible horizon, and the uncertainty in measuring those positions in the data is shown by the jitter of the points around the predicted path of the satellite. Satellite `x' has a similar path over the sky, and assuming the two satellites are passing with the same angular speeds at the same time, the predicted path of satellite `x' overlaps the six measured positions in the shaded region at the same time at satellite `y'. This means that $N = 20$ for satellite `y' candidates and $N = 6$ for satellite `x' candidates. It is important that the satellite `x' candidates are discarded as they are misidentifications, and none of the filters so far will remove them. In practice, if a satellite transmits hundreds of times during a pass, the number of additional candidates due to other satellite passes crossing over these can be quite large.

As another constraint, if $N < 4$, then these candidates would be discarded, as the algorithm performs best when $N$ is high. There was no evidence that certain satellites transmitted less frequently than this cadence over the course of a pass\footnote{The only caveat to this was Starlink satellites at 137.5 MHz which have been found to transmit every 100 s in \citep{grigg_starlink}. As these have already been reported in the literature, we omit these from the analyses in this paper for the 137.5 MHz dataset.}.

The predicted and measured positions of a satellite may follow a similar direction over the sky, but be offset in time, suggesting that these candidates need to be discarded. This was calculated by comparing the rate of change of the angle of travel of the satellite relative to the observer as a function of time, and the candidates were discarded if this exceeded a threshold.

It is therefore assumed that if the measured positions of a set of candidates have $N > 4$, have a similar path over the sky to its predicted positions, have a similar time incidence to its predicted positions, and have more candidates across the pass than any other coincident satellite, that these candidates are classed as identifications. These identifications are stored in a database with all accompanying metadata and the analysis on these is presented in Section \ref{results}.

\subsection{Estimating a Misidentification Rate}

Due to the enormous number of images and satellite trajectories in this analysis, it is inevitable that misidentifications are made. A misidentification is defined as either incorrectly labelling RFI as a satellite, or when one satellite is incorrectly identified as another. Estimating this percentage is important, so a simplified simulation was performed to better understand the occurrence of these misidentifications.

For this simulation, synthetic satellite signals were injected into the 229.7 MHz dataset described in Table \ref{tab:survey_table}. This dataset was chosen because it already contained known transmission from a number of satellites, and was an adequate length of 23 hours (${\approx}$42,000 images). This simulation ran completely independently from the results reported for the dataset in Section \ref{results}.

The parameters of the simulation were as follows:
\begin{itemize}
    \item 100 unique satellites (of the 18,540 with TLEs available under the constraints in Section \ref{making_detections}) were randomly chosen and checked to make sure they were not known transmitters;
    \item Each satellite was assigned a random flux density to be injected between 120 and 500 Jy/beam (this was not attenuated as a function of range or elevation); and
    \item An injection pattern was chosen for each satellite (only injected when a satellite's elevation was $>20^\circ$):
    \begin{itemize}
        \item every image;
        \item every three consecutive images then skipping the next three (and repeated); and
        \item for one image then skipping the next four (and repeated).
    \end{itemize}
\end{itemize}

Once these signals had been injected into the dataset, both the detection and identification algorithms from Sections \ref{making_detections} and \ref{making_identifications} were run over the data. Only the results for these 100 randomly selected satellites are reported here.

In total, 97 of the 100 randomly selected satellites reached an elevation of $>20^\circ$ within the dataset. For the XX polarisation, 219 of the 232 injected passes were recovered with a single misidentification. For the YY polarisation, 213 of the 232 injected passes were recovered, also with a single misidentification. Both misidentifications occurred on the same pass of the same satellite and were due to three STARLINK satellites having extremely similar predicted positions.

This simulation therefore estimates that the percentage of misidentifications is likely to be $<1\%$, with a ${\approx}$93\% recovery rate in the number of passes of satellites. Note that misidentifications are more likely at lower elevation satellite passes due to the slant orthographic projection used in images which has more degrees per pixel at lower elevations.

\subsection{Frequency Differencing Experiment}
\label{freq_dif}

Time differencing, although a successful technique for increasing the signal-to-noise ratio (SNR) of detections in images, has fundamental drawbacks. Image three in Figure \ref{fig:dif_images} shows the ISS in the current image (white) and the subtracted ISS from the previous frame (black). If the time between the subtracted images is reduced, these will begin to overlap at lower elevations where the angular speed of the satellite on the sky is slower. If the time between the subtracted images is increased, the sky moves more in this time and introduces more noise into the difference image. There are also more complex effects from photometric centroid displacement which are explained in \ref{app:sinusoidal}.

Satellite and FM radio transmissions are narrow-band in nature, compared to astrophysical sources of radio emission. If the satellite emission can be isolated in one frequency channel, this can be subtracted from a nearby frequency channel at the same time step which does not contain the satellite's emission. As the sky and the satellite now do not move, in theory, the SNR should be higher.

Frequency differencing was explored as an alternate method to time differencing. In this experiment, to ensure a fair comparison, both the time and frequency difference were done using fine channel (24.8 kHz) data. The frequency difference was performed by choosing a single fine channel where the satellite was visible, and subtracting from it another where the satellite was not visible from it for the same time step. The time difference was performed by choosing a single fine channel where the satellite is visible, and subtracting an image taken 40 s previously from it. This is illustrated in Figure \ref{fig:dif_images}.

Two scenarios were considered. The first was using the ISS with reflection of terrestrial FM transmission. In theory, the ISS should reflect transmission from the FM transmitter, and be visible at frequencies within its ${\approx}$0.1 MHz bandwidth. An image at the same time step, but at a frequency outside of the transmitter's bandwidth, can then be subtracted from this to do the frequency differencing.

The second scenario was using SOLRAD 7B as a known transmitter. From preliminary observations it was known that the emission received from SOLRAD 7B showed periodicity at 136.82 MHz over a narrow bandwidth which would be a suitable case for testing the frequency differencing approach for a transmitter. After quality control was conducted on the dataset, it was found that ORBCOMM FM 110 (NORAD 41182) was transmitting more brightly than SOLRAD 7B. The transmission from these two satellites overlapped, but ORBCOMM FM 110 has the highest received flux density at 136.85 MHz. This satellite was also included in the analysis.

To calculate the SNR of the satellite in each image, the fitted amplitude (A) in Equation \ref{eq:gaussian} is divided by an estimate of the noise of the image. To calculate the noise, the standard deviation ($\sigma$) of all pixels in the difference image is calculated after running two passes of removing all pixel intensities which are $>3\sigma$.

Each time step where an identification of the targeted satellite was made in both the time and frequency differenced images, the SNR of the identification was calculated. The ratio of the frequency difference SNR with the time difference SNR was averaged across all time steps and is presented in Section \ref{results_freq_dif}. This aimed to give an average ratio of the two SNRs across the entire pass. Note that this is tested on the XX polarisation images only.

\section{Results}
\label{results}

\begin{figure}[hbt!]
\centering
\includegraphics[width=0.99\linewidth]{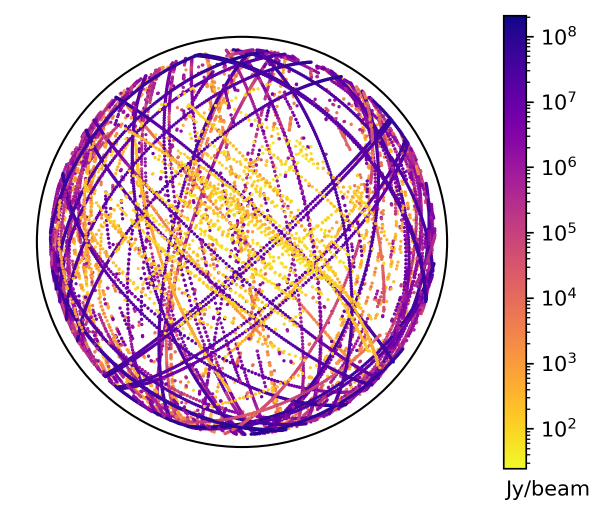}
\caption{A plot of the azimuth/elevation locations of all 29,005 individual identifications of satellites across all 18 datasets analysed in this work. The colourbar shows the logarithmic flux density of the identifications in Jy/beam. The black circle surrounding the detections shows the 0$^{\circ}$ elevation isoline.}
\label{fig:detections_on_sky_all}
\end{figure}

In total, 152 unique satellites were identified across these datasets. These results are displayed per dataset in Table \ref{tab:detections} and per NORAD I.D. in \ref{app:sat_table}. This comprised 14,165 and 14,840 individual identifications in the XX and YY polarisations, respectively. The closest identification of a satellite, from the distance predicted by the TLE, was BRICSAT (NORAD 44355) at 316 km, and the furthest was COSMOS 2453 (NORAD 35500) at 3,013 km. There was also the identification of an ex-geosynchronous  satellite EUTE 1-F5 (NORAD 19331) which was identified manually, and is the longest range identification of a satellite using these techniques.

Figure \ref{fig:detections_on_sky_all} shows the spatial spread of identifications over the sky. There is uniform coverage over the full sky, with identifications consistently being made down to the elevation cut-off of 20$^{\circ}$, showing that the identification algorithms are not biased to certain areas of the sky.

\begin{figure}[hbt!]
\centering
\includegraphics[width=0.99\linewidth]{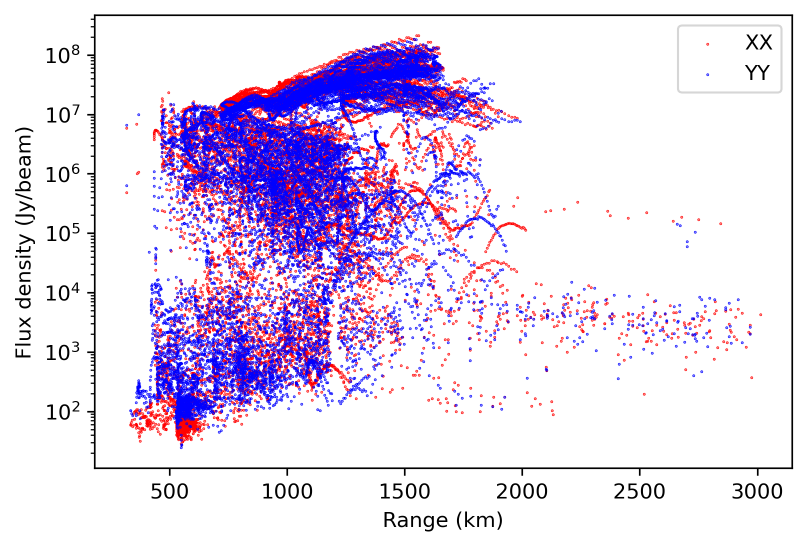}
\caption{The flux density and range of all 14,165 XX and 14,840 YY polarisation identifications of satellites across all 18 datasets analysed in this work.}
\label{fig:intensity_vs_range}
\end{figure}

The flux density of all identifications as a function of range is shown in Figure \ref{fig:intensity_vs_range}. 

The large majority of identifications were due to the direct detection of transmission from satellites. There was evidence of both IEMR and UEMR. Due to the high volume of identifications, only the most significant findings will be discussed, with the full results available in \ref{app:sat_table}. This Section summarises the results per dataset, and the implications of these results are discussed in Section \ref{discussion}.

\begin{table*}
\centering
\caption{Detections per dataset}
\label{tab:detections}
\begin{tabular}{|c|c|ccc|ccc|}
\hline
& & \multicolumn{3}{c|}{XX Polarisation} & \multicolumn{3}{c|}{YY polarisation} \\
\cline{3-8}
\multirow{-2}*{Frequency} & \multirow{-2}*{Duration} & Candidates & Identifications & Unique NORAD IDs & Candidates & Identifications & Unique NORAD IDs \\
\hline
\hline
73.4  & 37 min        & 0 & 0 & 0 & 0 & 0 & 0 \\
136.7 & 12 min        & 728 & 479 & 2 & 827 & 521 & 2 \\
137.5 & 19 h 35 min   & 17296 & 7088 & 71 & 19083 & 7154 & 73 \\
146.1 & 24 h 22 min   & 10573 & 3729 & 29 & 8623 & 4016 & 31 \\
150.8 & 4 min         & 0 & 0 & 0 & 0 & 0 & 0 \\
150.8 & 2 h 00 min    & 0 & 0 & 0 & 0 & 0 & 0 \\
159.4 & 132 h 35 min  & 14005 & 474 & 2 & 4633 & 666 & 2 \\
159.4 & 133 h 38 min  & 16738 & 734 & 3 & 9629 & 758 & 3 \\
159.4 & 22 h 57 min   & 5026 & 76 & 2 & 673 & 83 & 2 \\
160.2 & 55 h 04 min   & 8182 & 974 & 26 & 6023 & 994 & 25 \\
185.2 & 39 h 59 min   & 3272 & 90 & 1 & 1763 & 98 & 1 \\
229.7 & 23 h 14 min   & 3580 & 299 & 15 & 623 & 209 & 11 \\
324.2 & 4 h 11 min    & 1116 & 0 & 0 & 217 & 0 & 0 \\
324.2 & 2 h 37 min    & 31 & 0 & 0 & 10 & 0 & 0 \\
325.0 & 7 h 59 min    & 2160 & 0 & 0 & 792 & 0 & 0 \\
\hline
\end{tabular}


\end{table*}

\subsection{Known downlink frequencies}

The known downlink frequencies from this survey include the 146.1, 136.7, and 137.5 MHz frequency bands. All of these displayed a high number of identifications. Due to the widespread occurrence of high intensity signals, conducting radio astronomy at these frequencies would be exceedingly difficult.

\subsubsection{137.5 MHz}

The variations in this dataset were extreme compared to other datasets, with some image RMSs as low as 10 Jy/beam and as high as 10$^6$ Jy/beam. There are very short periods where no satellites are transmitting overhead and the Milky Way is clearly visible, but the majority of the time there could be one to 10 satellites visible. It was quite common to find a satellite transmitting so strongly that it would obscure emission from other transmitting satellites. For example, the identification of the ISS was only possible because nothing else was transmitting overhead at the same time.

At 137.5 MHz, the majority of satellites identified were ORBCOMM and SPACEBEE. There were also many identifications of STARLINK satellites in this dataset but not discussed in this work as they were reported in \citet{grigg_starlink}. Detected intensities were frequently on the order of 10$^7$ Jy/beam, with most satellites transmitting periodically.

The ex-geosynchronous satellite EUTE 1-F5 was identified serendipitously when manually assessing the quality of this dataset. This satellite was placed in a graveyard orbit in the year 2000. At a range of ${\approx}$36,000 km, it moves very slowly across the sky, meaning that time differencing will cause the emission to be completely subtracted. This particular satellite was transmitting strongly enough that its small movement across the sky was seen in the difference images and extremely obvious in the undifferenced images. It appeared to be rotating slowly as the flux density of the signal was sinusoidal over a period of multiple minutes (this was difficult to constrain due to constant high flux density signals from other satellites). This is expected to be easily identifiable when frequency differencing is used in the future.

The identification of PEGASUS R/B (NORAD 22491) was interesting, as it was not anticipated that a rocket body would be transmitting, especially one launched in 1993. It is unlikely to be reflecting terrestrial transmission as it was detected at a TLE predicted range of ${\approx}$1,500 km, slightly further than any previous detection of the ISS (the largest satellite in orbit) using this technique \citep{grigg}. It was manually verified for quality control and was the only TLE that matched up well for the period it was visible. \texttt{space-track.org} and \texttt{n2yo.com} both list the name of this satellite as PEGASUS R/B, but an amateur observational database \texttt{db.satnogs.org} lists the name as OXP1. On another amateur satellite website\footnote{\url{https://emitters.space/Emitters.html\#19}}, there is a note: "An unknown emitter on 137.05 MHz, initially reported by Raydel, has been identified by experts on the Hearsat list as OXP-1 that was launched in 1993. The search found that the closest match was for the Pegasus R/B but a comparison of radar cross section suggests the “correct assignments are 22491 for OXP-1 and 22489 for the Pegasus R/B stage".

Fifty four unique SPACEBEE satellites were identified in this dataset. Their received emissions were sporadic and consistently exceeded $10^{7}$ Jy/beam. With a longer acquisition time, there would likely be additional SPACEBEE satellites, as there are missing SPACEBEE satellites from the same launch. Identifications of SPACEBEE 63 and SPACEBEE 55 (NORADs 47442 and 47459 respectively) were interesting because they were launched two years earlier than most of the other identified SPACEBEE satellites. The reason why only two were identified when there were still others in orbit from the same launch is unknown.

\subsubsection{136.7 MHz}

In the short 12 minute dataset at 136.7 MHz, which was targeted on SOLRAD 7B, there were also many identifications of the satellite ORBCOMM 110. This satellite was also identified in the 137.5 MHz dataset. It is listed as transmitting at 137.288 MHz, so the mechanism behind the propagation of this transmission is unknown at this time. If this dataset were longer, there would have been more identifications made, as there were also many short bursts of radio energy which were difficult to identify without seeing the full passes of these satellites.

\begin{figure}[hbt!]
\centering
\includegraphics[width=1\linewidth]{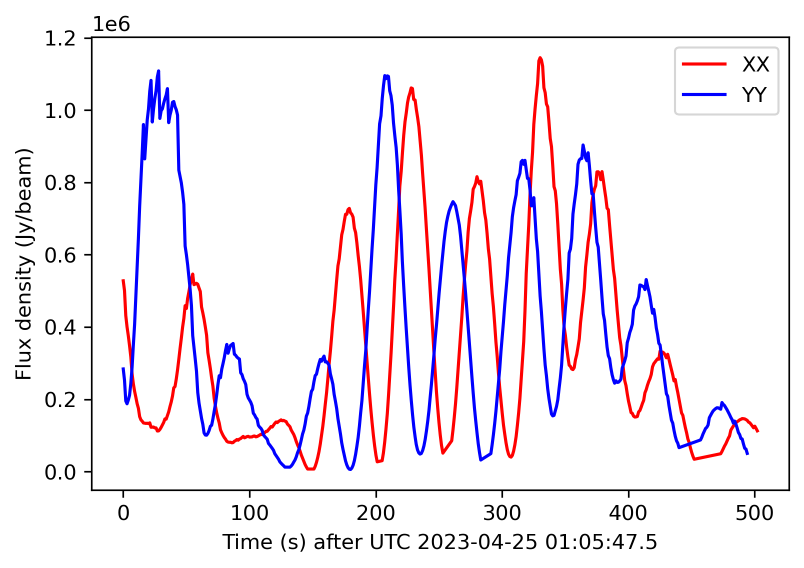}
\caption{The flux density of detections of SOLRAD 7B in the 136.7 MHz dataset. ORBCOMM FM110 is also transmitting strongly between 0 to 150 s.}
\label{fig:solrad_sinusoidal}
\end{figure}

Figure \ref{fig:solrad_sinusoidal} shows that the intensities of SOLRAD 7B in the two polarisations varied sinusoidally in time, out-of-phase by 90$^{\circ}$. This suggests that the satellite may no longer have stabilisation control and could be tumbling. More follow up observations would be needed to confirm this.

\subsubsection{146.1 MHz}
The quality of these data were high, with satellites being the main source of RFI. The majority of satellites identified at this frequency are cubesats or for amateur use, with a radar cross section of $<$1 m$^{2}$. All of the identified satellites at this frequency have their designated downlink frequency within the 0.926 MHz bandwidth of the acquisition. The cadence and measured flux density of the transmissions varied greatly, proving the robustness of the identification algorithms. Many of the cubesats had their transmission powers publicly available and all were less than 0.5 W in power.

BGUSAT, MAX VALIER SAT, and TRITON 1 are known to transmit broadband UEMR \citep{grigg}, and at this frequency they are all observed at their operational downlink frequency. The flux density of the received transmission is much higher at this frequency than any other frequency they have been observed at in this study and in \citet{grigg}. This could imply that the transmission detected at other frequencies is from a different mechanism and is likely UEMR. Unlike the UEMR from Starlink satellites \citep{grigg_starlink}, the transmission is visible for short multi-second bursts, and then disappears for short periods of time. For example, in general across all frequencies TRITON 1 disappears for longer periods than BGUSAT and MAX VALIER SAT.

OSCAR 11 (NORAD 14781) was launched in 1984 and is also an example of a long decommissioned satellite still emitting at radio frequencies. This satellite was identified over three passes and these were all times when the satellite was illuminated by the Sun, powering its battery. These were the only passes above 20$^{\circ}$ elevation during the span of the dataset.

HAMSAT (NORAD 28650) is yet another example of a satellite which has been decommissioned and yet is still transmitting. Its official website\footnote{\url{https://amsatindia.org/hamsat/}} states the reason for it being decommissioned as a "failure of on-board lithium batteries". There were two passes of the satellite during the dataset at 46$^{\circ}$ elevation, where it was illuminated by sunlight, and 34$^{\circ}$ elevation, where it was eclipsed by the Earth. It was identified in the first pass in the YY polarisation, and was not identified in the second pass, implying that the Sun may temporarily power the satellite, causing it to transmit.

\subsection{Protected frequencies}

The protected frequencies were 73.4, 150.8, 324.2, and 325 MHz. No satellites were identified in any of these datasets.

\subsubsection{324.2 \& 325 MHz}

The two datasets at 324.2 MHz and the other at 325 MHz were of high quality and had extremely low levels of RFI if any. The only high flux density signals in the datasets came from the Sun and its sidelobes.

\subsubsection{150.8 MHz}

There were two datasets acquired at 150.8 MHz. In the 150.8 MHz dataset acquired on 2023-06-13, there were two instances where an object appeared to move over the sky. These both followed the north/south common aircraft flight path over the telescope which is illustrated in Figure 8 in the paper by \citet{grigg}. There are some local mobile/defence transmitters across WA which have powers up to 50 W which could be causing these weak reflections. Follow up observations could be made to discern which transmitter the reflected transmission originated from, but is out of scope for this work.

Both datasets had 32, 28.9 kHz fine channels of data available, and although the main analysis processed these combined into a single 0.926 MHz channel, the fine channels were manually surveyed to determine if there was any narrowband emission in the data. It was noticed that there were some occasional broadband (covering all frequency channels) bursts, while others were narrow-band (one or two fine frequency channels). There was no apparent temporal association between any of these bursts.

\subsubsection{73.4 MHz}

Lastly, the 73.4 MHz dataset exhibited the least amount of RFI out of all the datasets in this work. There was no visible RFI in the 37 minutes of the dataset, which makes it ideal for other radio astronomy analyses.

\subsection{Comparison frequencies}

\subsubsection{229.7 MHz}

The 229.7 MHz dataset overlaps with the Australian digital television broadcasting band which extends up to 230 MHz. The Australian Radio Frequency Spectrum Plan\footnote{\url{https://www.acma.gov.au/sites/default/files/2021-07/Australian\%20Radiofrequency\%20Spectrum\%20Plan\%202021_Including\%20general\%20information.pdf}} lists the uses of this frequency for `broadcasting' (fixed and mobile) and aeronautical radionavigation. Overall the quality of this dataset was high, and the highest measured flux densities were from the Sun and Milky Way.

All identified satellites at this frequency were launched by the Commonwealth of Independent States (formerly USSR) and were of the COSMOS, GONETS, and STRELA series. All of these detected satellites displayed a very predictable pattern of transmitting within a single image, every 60s. This is why the identification counts are low for each satellite in Table \ref{tab:detections_by_norad}. An amateur satellite website\footnote{\url{https://www.orbitalfocus.uk/Frequencies/FrequenciesAll.php}} lists the reason for the transmission pattern as: "Message Signal every 60s, silent when receiving data message from ground".

\citet{soko_eda2} shows identifications at the same frequency a year and a half earlier. All of these identifications were also made in this work, apart from COSMOS 2438 (NORAD 32955). The EDA2 is most sensitive towards the zenith, and as the satellite was not detected during two high elevation passes, means it may have stopped transmitting during this year and a half period. It would be unlikely to be silent due to receiving data from the ground as it was travelling over Australia at this time. Our work additionally identifies some of the GONETS M series which are not identified in the work of \citet{soko_eda2} (GONETS M17 and 22 were launched after the \cite{soko_eda2} dataset).

Twenty three aircraft were also detected in this dataset at low intensities. 21 of these were on the common north/south flight path shown in Figure 6 in \citet{grigg}. They were mostly visible in the southern half of the sky, implying they are reflecting transmission closer to Perth (${\approx}$600 km south) where the aircraft would be taking off/landing. There was no evidence of any satellites reflecting this same terrestrial transmission.

\subsubsection{159.4 MHz}

There were three datasets used at this frequency. One was the ${\approx}$23 hour dataset acquired for this work on 2021-11-16 and the others were two ${\approx}$133 hour datasets acquired by \citet{soko_eda2} on 2020-06-26 with the AAVS2 and EDA2 acquiring data simultaneously.

The 2021-11-16 dataset yielded identifications of the satellites BGUSAT and TRITON 1, outside of these satellites' designated downlink frequency. This is another dataset showing detections of these two producing UEMR.

It was noticed in the study by \citet{soko_eda2} that there were many additional identifications of rocket bodies, debris objects, and geostationary orbit satellites. As none of these were identified in the 2021-11-16 dataset, there was the question of why these were being identified in the 2020-04-10 and 2020-06-26 datasets in the \citet{soko_eda2} work. The authors of Sokolowski et al. were kind enough to allow the two datasets from 2021-11-16 with the AAVS2 and EDA2 acquiring simultaneously for ${\approx}$133 hours to be made available for reprocessing with our new algorithms.

In our reprocessing of the data, for the polarisations XX/YY, BGUSAT was identified 316/486 times in the AAVS2 dataset and 469/515 times in the EDA2 dataset. \citet{soko_eda2} combines both polarisations into I (XX + YY) and does not differentiate between the datasets, but makes 499 identifications of BGUSAT. The timestamps for the two telescopes are on a very slightly different cadence, which makes combining the identification counts between the two datasets difficult, but there are more identifications of BGUSAT using our new algorithms.

Doing the same for TRITON 1, in our reprocessing of the data, for the polarisations XX/YY, TRITON 1 was identified in 158/180 images in the AAVS2 dataset and 240/237 images in the EDA2 dataset. \citet{soko_eda2} makes 191 identifications of TRITON 1. Again there are more identifications of the satellite using our new algorithms.

The only satellite which is identified less with the new algorithm is MAX VALIER SAT. There were six individual identifications only in the YY polarisation in the EDA2 dataset with our algorithm, whereas there were 17 individual identifications in \citet{soko_eda2}.

In the EDA2 dataset, SWIATOWID (NORADs 117) was additionally identified with our new algorithms.

There were no rocket bodies, debris objects, or geostationary orbit satellites identified with these new algorithms. This will be discussed in Section \ref{discussion}.

\subsection{Additional frequencies}
\subsubsection{160.2 MHz}
The 160.2 MHz dataset was recorded over 55 hours. There were 29 unique satellites identified over the two polarisations. Most of the identifications were Starlink satellites. The character of the emission is similar to the character reported at 159.4 MHz in the paper by \citet{grigg_starlink}. This emission is constant from image-to-image, unlike the transmission every 100 s that was exhibited at 137.5 MHz in the same study. The study also suggests that the emission is likely caused by the propulsion or avionics system, and is also likely to be broadband, meaning that it was classed as UEMR. 

This case is different. In the \citet{grigg_starlink} study, the identified Starlink satellites at 159.4 MHz are type `v1p5', whereas those identified in this dataset are all `v2\_mini'. All of these `v2\_mini' Starlink satellites were on one of four payloads which launched on 2023-02-27, 2023-04-19, 2023-05-19, and 2023-06-04. Our dataset was acquired on 2023-06-23 for just over two days. This means that the Starlink satellites that were launched on 2023-02-27 and 2023-04-19 were no longer being boosted to their operational altitude. Two examples can be seen for NORAD I.D.s 55695\footnote{\url{https://planet4589.org/space/con/star/sg76/S55695.jpg}} and 56301\footnote{\url{https://planet4589.org/space/con/star/sg81/S56301.jpg}} showing their historic altitudes. The propulsion system may still be engaged when these satellites are identified, but implies that this UEMR from Starlink satellites is likely to be ongoing when they arrive at their operational altitude.

BLUEWALKER 3 (NORAD 53807) was also identified at this frequency. In its technical specifications document submitted to the Federal Communications Commission, there is nothing mentioned about communications lower than 750 MHz\footnote{\url{https://www.fcc.report/ELS/AST-Science-LLC/1059-EX-CN-2020/265582.pdf}}. There was no evidence of reflection of terrestrial transmission off aircraft at this frequency, putting in question whether the detection of BLUEWALKER 3 was reflected energy. Its highest recorded flux density was in the YY polarisation at 1,043 Jy/beam. The source of this signal is currently unknown. 

The last two objects which were TRITON 1 and OBJECT D\footnote{The US Space Force typically assigns the designation "OBJECT" to unidentified satellites, using the letter that corresponds to the international designator letter of the launch}. TRITON 1 is a known broadband transmitter across many frequencies \citep{grigg}, but OBJECT D (either unnamed or unidentified) has not been previously identified, and very little information is available on it other than it was in a launch of other similarly named satellites from China on 2018-12-07.

\subsubsection{185.2 MHz}
The quality of this dataset was high in general. This being said, there were two transmitters on the southern horizon which were visible for the second half of the dataset (night time local time). Measured as a clockwise bearing angle from the vertical (north) in the images, the two transmitters are on a bearing of 182$^{\circ}$ and 195$^{\circ}$. Taking the transmitters from Table \ref{tab:transmitters}, the bearing angle on the images of the Morowa transmitter is 195$^{\circ}$ (${\approx}$300 km away) and the bearing of the two Perth transmitters is 185$^{\circ}$ (${\approx}$600 km away). These line up well with the transmitters seen on the horizon in the images. The effect of the two Perth transmitters interfering with each other can also be seen in the images.

The only satellite identified in this dataset was BGUSAT. There were also eleven aircraft detected, with most on a north/south or east/west trajectory. This dataset was acquired in February 2020, which was around the time that strict restrictions were put on International flights in Western Australia, which could explain the low detection rate of aircraft considering the dataset was 40 hours in length. Although the two Perth transmitters are 50 kW in power, comparable to transmitters in the FM band, the power is spread over a larger band-width (100 kHz for FM vs ${\approx}7$ MHz for DTV). This could explain the lack of satellite identifications from the reflection of this DTV transmission.

\subsection{Frequency Differencing}
\label{results_freq_dif}

\begin{table*}
\centering
\caption{Frequency differencing experiment results.}
\label{tab:freq_dif}
\begin{tabular}{|c|c|c|c|c|c|c|}
\hline
\begin{tabular}[c]{@{}c@{}}\textbf{Frequency} \\ \textbf{with signal}\end{tabular}
& \begin{tabular}[c]{@{}c@{}}\textbf{Frequency} \\ \textbf{with no signal}\end{tabular}
& \begin{tabular}[c]{@{}c@{}}\textbf{Target} \\ \textbf{NORAD I.D.}\end{tabular}
& \textbf{Target name}
& \begin{tabular}[c]{@{}c@{}}\textbf{Highest} \\ \textbf{elevation ($^\circ$)}\end{tabular}
& \begin{tabular}[c]{@{}c@{}}\textbf{Average} \\ \textbf{SNR increase}\end{tabular} \\
\hline
\multicolumn{6}{c}{Reflection} \\
\hline
96.92   & 97.06   & 25544 & ISS              & 71 & 0.86  \\
96.92   & 97.06   & 25544 & ISS              & 76 & 1.15  \\
98.60   & 98.28   & 25544 & ISS              & 89 & 1.39  \\
\hline
\multicolumn{6}{c}{Transmission} \\
\hline
136.82   & 136.39 & 1291  & SOLRAD 7B        & 74 & 2.23  \\
136.85   & 137.00 & 41182 & ORBCOMM FM 110   & 54 & 1.55  \\
\hline
\end{tabular}
\end{table*}

Table \ref{tab:freq_dif} summarises the results from the four datasets which were used to test the frequency differencing technique.

\subsubsection{Reflected FM transmission}
The three passes of the ISS at the two different frequencies showed between a 39\% increase to an 11\% decrease in SNR of the ISS. At these FM frequencies, there were no astronomical sources visible in the images, which meant that the noise was dominated by sidelobes and the FM transmitters visible on the horizon. In some cases in the time difference, the direct transmission from these FM transmitters was partially subtracted as it was visible in both images used in the differencing. For frequency differencing, the direct transmission would never be subtracted as it was only visible in one of the images used in differencing by design. This explains why there was a 11\% decrease in the SNR of the ISS in the frequency differenced images.

This being said, frequency differencing increases the average SNR of detections by 39\% and 15\% in the other two datasets. This was more due to a reduction in the noise than an increase in the signal used to calculate the SNR.

\subsubsection{Transmission}
For the pass of SOLRAD 7B, frequency differencing increased the average SNR by 123\%, and the pass of ORBCOMM FM 110 by 55\%. The reason for such a different result for both satellites is that the intensities of SOLRAD 7B oscillate as shown in Figure \ref{fig:solrad_sinusoidal}. This means that SOLRAD 7B's measured flux density will be low in some images, but the subtracted flux density from the image 40 s prior will be high, resulting in a low SNR in the time difference.

The lack of high flux density transmitters on the horizon, along with astronomical sources of radio emission being slightly visible in the images results in the frequency difference increasing the SNR of the two satellites. This will be the favoured approach for detecting direct transmission from satellites in the future.

\section{Discussion}
\label{discussion}

The results presented in this work demonstrate significant improvements in our algorithms for the detection and identification of satellites. Imaging the whole sky presents advantages in that multiple satellites can be characterised simultaneously in the same image, providing rich datasets worthy of many types of astrophysical analysis. 

A total of 152 unique satellites were identified across frequencies over a 250 MHz range. The 18 datasets analysed varied considerably in the type of signals received by the telescope. Different sources of radio energy such as terrestrial transmitters, meteors, the Sun, astrophysical sources of broadband synchrotron emission, aircraft, lightning, and other anthropogenic sources were handled to robustly identify satellites. Different satellite transmitting cadences were handled by the algorithm to identify satellites which transmitted sporadically over small portions of their passes. 

There is a slight difference in the total number of identifications in each polarisation, with 14,165 individual identifications in the XX polarisation and 14,840 in YY. This is more skewed in some individual datasets shown in Table \ref{tab:detections}. This can be influenced by the antenna beam response in each polarisation, as well as physical properties of the satellite itself (polarisation characteristic and orientation of the satellite's antenna). For example, the 229.7 MHz dataset has ${\approx}$43\% more identifications in the XX polarisation. All of the satellites identified at 229.7 MHz followed a north/south path over the sky, meaning the satellite will have slower apparent angular speed in the north and south of the sky where the XX polarisation is more sensitive. This can explain the difference observed between the two polarisations. An analysis of how the antenna's physical characteristics influences detection rates would be interesting but is out of scope for this work.

The majority of identifications were in LEO as shown in Table \ref{tab:detections_by_norad}. For example, identifications at 229.7 MHz show that medium earth orbit (MEO) satellites can also routinely be identified. Since the majority of satellites are in LEO, being able to identify satellites beyond LEO shows good promise for future broader frequency surveys. The manual identification of the (now in graveyard orbit) geosynchronous satellite EUTE 1-F5 shows that satellites with a range of over 36,000 km can be detected. Tests show that it could repeatedly be detected in non-differenced images of the sky, showing that this technique could be applied to geosynchronous/geostationary range searches in the future.

Figure \ref{fig:intensity_vs_range} shows that the measured flux density of satellite transmission can regularly exceed 10$^6$ Jy/beam, with many higher than 10$^8$ Jy/beam. The 137.5 MHz frequency band is host to many of these satellites, but is also widespread among other frequency bands. This indicates challenges that will be faced in attempts to make astrophysical observations at these frequencies. Satellite constellations such as ORBCOMM and SPACEBEE are legally allowed to transmit at these frequencies, but their transmission obscures astrophysical signals over the whole sky in most instances.

\subsection{Comparison of algorithms from \citet{soko_eda2}}
Reprocessing of the two five and a half day datasets from the work of \citet{soko_eda2} shows a marked decrease in the number of misidentifications. Our work introduces an additional final step in the identification algorithm that takes candidate detections and ensures they follow the expected trajectory of the satellite over the sky, which eliminates many of these misidentifications. This new algorithm also identifies more of the same objects, showing that they are robust over a long observing period. The simulation run to estimate a misidentification rate also shows an extremely low occurrence of these.

\subsection{Incomplete database information}
At the time of writing this paper, \texttt{space-track.org} tracks ${\approx}$45,000 objects in Earth orbit. The maintenance and validation of this database is an enormous effort and is in-part reliant on satellite operators being able to identify their own spacecraft. Our detection of PEGASUS R/B appearing to `transmit', along with the information from the amateur satellite page showing that it was likely OXP-1, is an example of inconsistencies that could appear in this database. There were also at least ten passes of objects that travelled across the sky at angular speeds of LEO satellites, which could not be attributed to anything in the TLE list obtained from \texttt{space-track.org}.

\subsection{Satellite behaviour}
The lack of identifications in any of the protected frequency bands is encouraging. The degree of protection over the 73.4, 150.8, 324.2, and 325.0 MHz frequencies varies for each, but it is promising that in these datasets there are no identifications made. These results can serve as a baseline for future studies to ensure satellites are transmitting safely and correctly.

\subsection{Satellite operational status}
A number of satellites were identified which have been officially decommissioned, and yet are still emitting at radio frequencies. There could be a couple of reasons for this. Although decommissioned, these satellites may still be performing a limited subset of functions, or loss of control of the satellite may mean that they will keep transmitting until they lose power or ultimately decay. For the three listed examples of SOLRAD 7B, HAMSAT, and OSCAR 11, they all display the following characteristics: they were all launched some time ago (1965, 2005, and 1984 respectively), are publicly listed as decommissioned, all are illuminated by the Sun when detected in the data (the Sun potentially temporarily charges up the battery), and they all show sinusoidal variation in their lightcurves (potentially due to lack of stabilisation control). 

Satellites that remain active after decommissioning need to continue to be tracked and monitored, but are shown to sometimes continue to take up an allocation in an already congested radio allocation zone. The legislative landscape and permit allocation procedures have changed significantly since some of these satellites were launched and will continue to change in the future. There needs to be stronger enforcement of procedures for the decommissioning of satellites so they do not continue to be a source of RFI for radio astronomy.

\subsection{UEMR}
In the work by \citet{grigg_starlink}, the authors quote a discussion with SpaceX (the company operating the STARLINK constellation), in which SpaceX engineers propose that the propulsion or avionics system is the mechanism behind the likely broadband UEMR detected at 159.4 MHz. In this particular example, the STARLINK satellites were being boosted to their operational altitude. Our new results show that the STARLINK satellites are also transmitting at 160.2 MHz when they are already at their operational altitude. Multiple passes of the same STARLINK satellites are also identified, bringing into question whether the propulsion system is the cause of the observed signal. 

None of the model `v1p5' Starlink satellites which were detected at 159.4 MHz in \citet{grigg_starlink} were detected at 160.2 MHz where we detect the `v2\_mini' model Starlink satellites in this work. For the `v2\_mini' Starlink satellites which were identified above 85$^{\circ}$ elevation, the average flux density was was 78 Jy/beam at a range of 532 km (EIRP $\approx$ 2.6 $\mu$W). The work by \citet{starlink_bassa} confirms that the types of UEMR transmitted by the version 1 and version 2 are at different frequencies, and that the version 2 (mini and mini direct-to-cell) satellites were detected on average 32 times brighter than the version 1.0 satellites. Further monitoring of the STARLINK satellites will be crucial for assessing the impact of their UEMR on radio astronomy.

In addition to STARLINK, there were other satellites also transmitting broadband UEMR. BGUSAT, MAX VALIER SAT, and TRITON 1 continue to show evidence of being detected in many frequency channels. At the date of writing this paper, BGUSAT has decayed, but MAX VALIER SAT and TRITON 1 will continue to be sources of interference for radio astronomy research. BLUEWALKER 3 also shows evidence of an unknown source of radio frequency transmission at 160.2 MHz, which is not listed as a downlink frequency for the satellite. An in-depth investigation of whether satellites are transmitting outside of their designated downlink frequency bands is beyond the scope for this work, but Table \ref{tab:detections_by_norad} provides detailed information on the identification of individual satellites that interested readers may use to assess this further.

\subsection{Aircraft}
Aircraft prove to be an ultra-bright source of RFI at frequencies of high powered terrestrial transmitters, and this has been characterised for the FM band in \citet{grigg}. The datasets at frequencies of 185.2 and 229.7 both show evidence of aircraft, although at lower intensities than in the FM band. This again showcases the potential to use bi-static radar in the form of a radio telescope to passively search for and locate aircraft. With larger baselines, it would be possible to perform accurate parallax measurements of aircraft to also determine range.

\subsection{Frequency differencing}
Testing of the frequency differencing method has shown that in general, the SNR of satellite detections is increased compared to time differencing. The caveat to this is that time differencing may give a better SNR when there are high measured flux densities from terrestrial transmitters visible on the horizon. Frequency differencing additionally minimises the impact from broadband sources of radio emission, and removes the subtracted satellite emission shown in image 3 in Figure \ref{fig:dif_images}. Ultimately, frequency differencing will be used in future all-sky surveys for the characterisation of satellites with the EDA2.

\subsection{Future work and improvements}
These algorithms are now ready to be deployed in an automated manner over a more targeted survey on key science frequencies of the SKA-Low observing bandwidth of 50 - 350 MHz. With the addition of 32 channels of 28.9 kHz bandwidth data, such a study would provide very detailed information about the implications of satellites on radio astronomy observations at these frequencies. This will be the next work in this series, and will form an enhanced baseline for satellite activity, which can be repeated in the future to assess the change in radio frequency environment. 

Although the predicted misidentification rate is $<1\%$, this may not hold for datasets with a high number of visible satellites or high levels of RFI. As frequency differencing will be the favoured algorithm for processing the next larger survey, a more comprehensive simulation will be performed in that work.

Time differencing still holds value, for if a satellite produces broadband UEMR, then it is unlikely to be detected with frequency differencing. A potential improvement could be to decrease the time between the two images used in the time difference for higher frequencies as the size of the PSF in degrees decreases at higher frequencies.

Future work will likely attempt to expand on the polarisation analysis, looking to compare the observed polarisation response to physical antenna characteristics. Using far longer integration times to do time differencing may prove useful for detecting satellites in orbits out to MEO and geostationary. Building a database of unidentifiable detections will also be useful for future orbit determination analysis. TLEs could then be derived for these objects.

Although transmission bandwidth of some satellites is known, currently it is difficult to determine if this leaks into adjacent smaller frequency channels when the current observing frequency bandwidth is 0.926 MHz. There is also time variability of the received transmission which is difficult to model. The future survey will have finer frequency resolution across multiple channels which will enable a more detailed study of transmission bandwidths.

Lastly, the identifications made in this study have been stored in a database for future use. Future identifications will be added to this database, which may reveal temporal trends over the coming years. This can be used in conjunction with conventional approaches to space situational awareness such as radar and optical telescopes.

\section{Conclusion}
\label{conclusion}

Using all-sky imaging to comprehensively characterise and identify satellites across the SKA-Low science frequencies with two orthogonal polarisations has made critical insights into the behaviour of satellites at these low frequencies. This study offers evidence to the challenges that future low frequency science will face from the exponentially growing population of satellites in Earth orbit. Examples are also shown of satellites transmitting UEMR and so called `zombie satellites' transmitting after their decommissioning or demise. The identification algorithms put forward in this work are ready to be deployed in a systematic and targeted survey of the SKA-Low observing frequencies to assess the widespread effects of satellite transmission. LEO and MEO automated detection and identification is currently achievable, with a detection in a graveyard geosynchronous orbit paving the way for future algorithmic development. Ultimately, we will develop comprehensive tools for space situational awareness, offering unique scientific insight into the behaviour of satellite operations.

\begin{acknowledgement}
We would like to acknowledge the Wajarri Yamatji people as the traditional custodians of the Murchison Radio Observatory site, Inyarrimanha Ilgardi Bundara \citep{5164979}. The data processing for this work was run at the DUG Technology Perth HPC centre. We also wish to thank the developers of several key packages to this work, namely \texttt{MIRIAD} \citep{miriad} for the preprocessing, and the Python packages \texttt{Skyfield} \citep{skyfield}, \texttt{Astropy} \citep{astropy}, and \texttt{Numpy} \citep{numpy}.
\end{acknowledgement}

\printendnotes

\printbibliography

@article{soko_eda2, 
    title={A Southern-Hemisphere all-sky radio transient monitor for SKA-Low prototype stations},
    volume={38}, 
    DOI={10.1017/pasa.2021.16}, 
    journal={Publications of the Astronomical Society of Australia}, 
    publisher={Cambridge University Press}, 
    author={Sokolowski, M. and Wayth, R. B. and Bhat, N. D. R. and Price, D. and Broderick, J. W. and Bernardi, G. and Bolli, P. and Chiello, R. and Comoretto, G. and Crosse, B. and et al.}, 
    year={2021}, 
    pages={e023}
}

@article{grigg,
    title = {DUG Insight: A software package for big-data analysis and visualisation, and its demonstration for passive radar space situational awareness using radio telescopes},
    journal = {Astronomy and Computing},
    volume = {40},
    pages = {100619},
    year = {2022},
    issn = {2213-1337},
    doi = {10.1016/j.ascom.2022.100619},
    url = {https://www.sciencedirect.com/science/article/pii/S2213133722000452},
    author = {D. Grigg and S.J. Tingay and M. Sokolowski and R.B. Wayth}
}

@inproceedings{aavs2,
    author = {A. J. J. van Es and M. G.  Labate and M. F.  Waterson and J. Monari and P. Bolli and D. Davidson and R. Wayth and M. Sokolowski and P. Di Ninni and G. Pupillo and G. Macario and G. Virone and L. Ciorba and F. Paonessa},
    title = {{A prototype model for evaluating SKA-LOW station calibration}},
    volume = {11445},
    booktitle = {Ground-based and Airborne Telescopes VIII},
    editor = {Heather K. Marshall and Jason Spyromilio and Tomonori Usuda},
    organization = {International Society for Optics and Photonics},
    publisher = {SPIE},
    pages = {1449 -- 1468},
    year = {2020},
    doi = {10.1117/12.2562391},
    URL = {https://doi.org/10.1117/12.2562391}
}

@misc{eda2,
  doi = {10.48550/ARXIV.2112.00908},
  url = {https://arxiv.org/abs/2112.00908},
  author = {Wayth, Randall and Sokolowski, Marcin and Broderick, Jess and Tingay, Steven J. and Bhushan, Raunaq and Booler, Tom and Chiello, Riccardo and Davidson, David B. and Emrich, David and Juswardy, Budi and Kenney, David and Macario, Giulia and Magro, Alessio and Mattana, Andrea and Minchin, David and Monari, Jader and McPhail, Andrew and Perini, Federico and Pupillo, Giuseppe and Schiaffino, Marco and Subrahmanya, Ravi and van Es, Andre and Walker, Mia and Waterson, Mark},
  title = {The Engineering Development Array 2: design, performance and lessons from an SKA-Low prototype station},
  publisher = {arXiv},
  year = {2021},
  copyright = {arXiv.org perpetual, non-exclusive license}
}

@INPROCEEDINGS{miriad,
       author = {{Sault}, R.~J. and {Teuben}, P.~J. and {Wright}, M.~C.~H.},
        title = "{A Retrospective View of MIRIAD}",
    booktitle = {Astronomical Data Analysis Software and Systems IV},
         year = 1995,
       editor = {{Shaw}, R.~A. and {Payne}, H.~E. and {Hayes}, J.~J.~E.},
       series = {Astronomical Society of the Pacific Conference Series},
       volume = {77},
        month = jan,
        pages = {433},
archivePrefix = {arXiv},
       eprint = {astro-ph/0612759},
 primaryClass = {astro-ph},
       adsurl = {https://ui.adsabs.harvard.edu/abs/1995ASPC...77..433S}
}

@ARTICLE{tingay_eda_ssa,
       author = {{Tingay}, S.~J. and {Sokolowski}, M. and {Wayth}, R. and {Ung}, D.},
        title = "{A survey of spatially and temporally resolved radio frequency interference in the FM band at the Murchison Radio-astronomy Observatory}",
      journal = {\pasa},
         year = 2020,
        month = sep,
       volume = {37},
          eid = {e039},
        pages = {e039},
          doi = {10.1017/pasa.2020.32},
archivePrefix = {arXiv},
       eprint = {2008.05918},
 primaryClass = {astro-ph.IM},
       adsurl = {https://ui.adsabs.harvard.edu/abs/2020PASA...37...39T}
}

@MISC{skyfield,
       author = {{Rhodes}, Brandon},
        title = "{Skyfield: High precision research-grade positions for planets and Earth satellites generator}",
         year = 2019,
        month = jul,
          eid = {ascl:1907.024},
        pages = {ascl:1907.024},
archivePrefix = {ascl},
       eprint = {1907.024},
       adsurl = {https://ui.adsabs.harvard.edu/abs/2019ascl.soft07024R}
}

@ARTICLE{steve,
       author = {{Prabu}, Steve and {Hancock}, P. and {Zhang}, X. and {Tingay}, S.~J.},
        title = "{A low-frequency blind survey of the low Earth orbit environment using non-coherent passive radar with the Murchison widefield array}",
      journal = {\pasa},
         year = 2020,
        month = dec,
       volume = {37},
          eid = {e052},
        pages = {e052},
          doi = {10.1017/pasa.2020.40},
archivePrefix = {arXiv},
       eprint = {2006.04327},
 primaryClass = {astro-ph.IM},
       adsurl = {https://ui.adsabs.harvard.edu/abs/2020PASA...37...52P}
}

@article{lofar_desc,
	author = {{van Haarlem, M. P.} and {Wise, M. W.} and {Gunst, A. W.} and {Heald, G.} and {McKean, J. P.} and {Hessels, J. W. T.} and {de Bruyn, A. G.} and {Nijboer, R.} and {Swinbank, J.} and {Fallows, R.} and {Brentjens, M.} and {Nelles, A.} and {Beck, R.} and {Falcke, H.} and {Fender, R.} and {H\"orandel, J.} and {Koopmans, L. V. E.} and {Mann, G.} and {Miley, G.} and {R\"ottgering, H.} and {Stappers, B. W.} and {Wijers, R. A. M. J.} and {Zaroubi, S.} and {van den Akker, M.} and {Alexov, A.} and {Anderson, J.} and {Anderson, K.} and {van Ardenne, A.} and {Arts, M.} and {Asgekar, A.} and {Avruch, I. M.} and {Batejat, F.} and {B\"ahren, L.} and {Bell, M. E.} and {Bell, M. R.} and {van Bemmel, I.} and {Bennema, P.} and {Bentum, M. J.} and {Bernardi, G.} and {Best, P.} and {B\^{\i}rzan, L.} and {Bonafede, A.} and {Boonstra, A.-J.} and {Braun, R.} and {Bregman, J.} and {Breitling, F.} and {van de Brink, R. H.} and {Broderick, J.} and {Broekema, P. C.} and {Brouw, W. N.} and {Br\"uggen, M.} and {Butcher, H. R.} and {van Cappellen, W.} and {Ciardi, B.} and {Coenen, T.} and {Conway, J.} and {Coolen, A.} and {Corstanje, A.} and {Damstra, S.} and {Davies, O.} and {Deller, A. T.} and {Dettmar, R.-J.} and {van Diepen, G.} and {Dijkstra, K.} and {Donker, P.} and {Doorduin, A.} and {Dromer, J.} and {Drost, M.} and {van Duin, A.} and {Eisl\"offel, J.} and {van Enst, J.} and {Ferrari, C.} and {Frieswijk, W.} and {Gankema, H.} and {Garrett, M. A.} and {de Gasperin, F.} and {Gerbers, M.} and {de Geus, E.} and {Grie\ss{}meier, J.-M.} and {Grit, T.} and {Gruppen, P.} and {Hamaker, J. P.} and {Hassall, T.} and {Hoeft, M.} and {Holties, H. A.} and {Horneffer, A.} and {van der Horst, A.} and {van Houwelingen, A.} and {Huijgen, A.} and {Iacobelli, M.} and {Intema, H.} and {Jackson, N.} and {Jelic, V.} and {de Jong, A.} and {Juette, E.} and {Kant, D.} and {Karastergiou, A.} and {Koers, A.} and {Kollen, H.} and {Kondratiev, V. I.} and {Kooistra, E.} and {Koopman, Y.} and {Koster, A.} and {Kuniyoshi, M.} and {Kramer, M.} and {Kuper, G.} and {Lambropoulos, P.} and {Law, C.} and {van Leeuwen, J.} and {Lemaitre, J.} and {Loose, M.} and {Maat, P.} and {Macario, G.} and {Markoff, S.} and {Masters, J.} and {McFadden, R. A.} and {McKay-Bukowski, D.} and {Meijering, H.} and {Meulman, H.} and {Mevius, M.} and {Middelberg, E.} and {Millenaar, R.} and {Miller-Jones, J. C. A.} and {Mohan, R. N.} and {Mol, J. D.} and {Morawietz, J.} and {Morganti, R.} and {Mulcahy, D. D.} and {Mulder, E.} and {Munk, H.} and {Nieuwenhuis, L.} and {van Nieuwpoort, R.} and {Noordam, J. E.} and {Norden, M.} and {Noutsos, A.} and {Offringa, A. R.} and {Olofsson, H.} and {Omar, A.} and {Orr\'u, E.} and {Overeem, R.} and {Paas, H.} and {Pandey-Pommier, M.} and {Pandey, V. N.} and {Pizzo, R.} and {Polatidis, A.} and {Rafferty, D.} and {Rawlings, S.} and {Reich, W.} and {de Reijer, J.-P.} and {Reitsma, J.} and {Renting, G. A.} and {Riemers, P.} and {Rol, E.} and {Romein, J. W.} and {Roosjen, J.} and {Ruiter, M.} and {Scaife, A.} and {van der Schaaf, K.} and {Scheers, B.} and {Schellart, P.} and {Schoenmakers, A.} and {Schoonderbeek, G.} and {Serylak, M.} and {Shulevski, A.} and {Sluman, J.} and {Smirnov, O.} and {Sobey, C.} and {Spreeuw, H.} and {Steinmetz, M.} and {Sterks, C. G. M.} and {Stiepel, H.-J.} and {Stuurwold, K.} and {Tagger, M.} and {Tang, Y.} and {Tasse, C.} and {Thomas, I.} and {Thoudam, S.} and {Toribio, M. C.} and {van der Tol, B.} and {Usov, O.} and {van Veelen, M.} and {van der Veen, A.-J.} and {ter Veen, S.} and {Verbiest, J. P. W.} and {Vermeulen, R.} and {Vermaas, N.} and {Vocks, C.} and {Vogt, C.} and {de Vos, M.} and {van der Wal, E.} and {van Weeren, R.} and {Weggemans, H.} and {Weltevrede, P.} and {White, S.} and {Wijnholds, S. J.} and {Wilhelmsson, T.} and {Wucknitz, O.} and {Yatawatta, S.} and {Zarka, P.} and {Zensus, A.} and {van Zwieten, J.}},
	title = {LOFAR: The LOw-Frequency ARray},
	DOI= "10.1051/0004-6361/201220873",
	url= "https://doi.org/10.1051/0004-6361/201220873",
	journal = {A\&A},
	year = 2013,
	volume = 556,
	pages = "A2",
	month = "",
}

@ARTICLE{hennessy,
       author = {{Hennessy}, Brendan and {Rutten}, Mark and {Young}, Robert and {Tingay}, Steven and {Summers}, Ashley and {Gustainis}, Daniel and {Crosse}, Brian and {Sokolowski}, Marcin},
        title = "{Establishing the Capabilities of the Murchison Widefield Array as a Passive Radar for the Surveillance of Space}",
      journal = {Remote Sensing},
         year = 2022,
        month = may,
       volume = {14},
       number = {11},
        pages = {2571},
          doi = {10.3390/rs14112571},
archivePrefix = {arXiv},
       eprint = {2206.02357},
 primaryClass = {eess.SP},
       adsurl = {https://ui.adsabs.harvard.edu/abs/2022RemS...14.2571H}
}

@ARTICLE{grigg_starlink,
       author = {{Grigg}, D. and {Tingay}, S.~J. and {Sokolowski}, M. and {Wayth}, R.~B. and {Indermuehle}, B. and {Prabu}, S.},
        title = "{Detection of intended and unintended emissions from Starlink satellites in the SKA-Low frequency range, at the SKA-Low site, with an SKA-Low station analogue}",
      journal = {\aap},
         year = 2023,
        month = oct,
       volume = {678},
          eid = {L6},
        pages = {L6},
          doi = {10.1051/0004-6361/202347654},
archivePrefix = {arXiv},
       eprint = {2309.15672},
 primaryClass = {astro-ph.IM},
       adsurl = {https://ui.adsabs.harvard.edu/abs/2023A&A...678L...6G}
}

@ARTICLE{di_vruno,
       author = {{Di Vruno}, F. and {Winkel}, B. and {Bassa}, C.~G. and {J{\'o}zsa}, G.~I.~G. and {Brentjens}, M.~A. and {Jessner}, A. and {Garrington}, S.},
        title = "{Unintended electromagnetic radiation from Starlink satellites detected with LOFAR between 110 and 188 MHz}",
      journal = {\aap},
         year = 2023,
        month = aug,
       volume = {676},
          eid = {A75},
        pages = {A75},
          doi = {10.1051/0004-6361/202346374},
archivePrefix = {arXiv},
       eprint = {2307.02316},
 primaryClass = {astro-ph.IM},
       adsurl = {https://ui.adsabs.harvard.edu/abs/2023A&A...676A..75D}
}

@ARTICLE{scipy,
  author  = {Virtanen, Pauli and Gommers, Ralf and Oliphant, Travis E. and
            Haberland, Matt and Reddy, Tyler and Cournapeau, David and
            Burovski, Evgeni and Peterson, Pearu and Weckesser, Warren and
            Bright, Jonathan and {van der Walt}, St{\'e}fan J. and
            Brett, Matthew and Wilson, Joshua and Millman, K. Jarrod and
            Mayorov, Nikolay and Nelson, Andrew R. J. and Jones, Eric and
            Kern, Robert and Larson, Eric and Carey, C J and
            Polat, {\.I}lhan and Feng, Yu and Moore, Eric W. and
            {VanderPlas}, Jake and Laxalde, Denis and Perktold, Josef and
            Cimrman, Robert and Henriksen, Ian and Quintero, E. A. and
            Harris, Charles R. and Archibald, Anne M. and
            Ribeiro, Ant{\^o}nio H. and Pedregosa, Fabian and
            {van Mulbregt}, Paul and {SciPy 1.0 Contributors}},
  title   = {{{SciPy} 1.0: Fundamental Algorithms for Scientific
            Computing in Python}},
  journal = {Nature Methods},
  year    = {2020},
  volume  = {17},
  pages   = {261--272},
  adsurl  = {https://rdcu.be/b08Wh},
  doi     = {10.1038/s41592-019-0686-2},
}

@ARTICLE{eor_explanation,
       author = {{Morales}, Miguel F. and {Wyithe}, J. Stuart B.},
        title = "{Reionization and Cosmology with 21-cm Fluctuations}",
      journal = {\araa},
         year = 2010,
        month = sep,
       volume = {48},
        pages = {127-171},
          doi = {10.1146/annurev-astro-081309-130936},
archivePrefix = {arXiv},
       eprint = {0910.3010},
 primaryClass = {astro-ph.CO},
       adsurl = {https://ui.adsabs.harvard.edu/abs/2010ARA&A..48..127M}
}

@ARTICLE{radio_dynamic_zones,
  author={Zheleva, Mariya and Anderson, Christopher R. and Aksoy, Mustafa and Johnson, Joel T. and Affinnih, Habib and DePree, Christopher G.},
  journal={IEEE Communications Magazine}, 
  title={Radio Dynamic Zones: Motivations, Challenges, and Opportunities to Catalyze Spectrum Coexistence}, 
  year={2023},
  volume={61},
  number={6},
  pages={156-162},
  doi={10.1109/MCOM.005.2200389}}

@article{rfi_gnss,
author = {Gilloire, André and Sizun, Hervé},
year = {2009},
month = {10},
pages = {625-638},
title = {RFI mitigation of GNSS signals for radio astronomy: Problems and current techniques},
volume = {64},
journal = {Annales des Télécommunications},
doi = {10.1007/s12243-009-0112-3}
}

@article{stevep_2,
author = {Prabu, Steve and Hancock, Paul and Zhang, Xiang and Tingay, Steven},
year = {2020},
month = {03},
pages = {},
title = {The development of non-coherent passive radar techniques for space situational awareness with the Murchison Widefield Array},
volume = {37},
journal = {Publications of the Astronomical Society of Australia},
doi = {10.1017/pasa.2020.1}
}

@article{tle_uncert,
author = {Geul, Jacco and Mooij, Erwin and Noomen, Ron},
year = {2017},
month = {03},
pages = {},
title = {TLE Uncertainty Estimation using Robust Weighted Differencing},
journal = {Advances in Space Research},
doi = {10.1016/j.asr.2017.02.038}
}

@INPROCEEDINGS{rqz,
  author={Wilson, C. and Chow, K. and Harvey-Smith, L. and Indermuehle, B. and Sokolowski, M. and Wayth, R.},
  booktitle={2016 International Conference on Electromagnetics in Advanced Applications (ICEAA)}, 
  title={The Australian Radio Quiet Zone — Western Australia: Objectives, implementation and early measurements}, 
  year={2016},
  volume={},
  number={},
  pages={922-923},
  doi={10.1109/ICEAA.2016.7731554}}

@book{passive_radar,
title = {Chapter 16 - Passive Bistatic Radar},
editor = {Nicholas D. Sidiropoulos and Fulvio Gini and Rama Chellappa and Sergios Theodoridis},
series = {Academic Press Library in Signal Processing},
publisher = {Elsevier},
volume = {2},
pages = {813-855},
year = {2014},
booktitle = {Academic Press Library in Signal Processing: Volume 2},
issn = {2351-9819},
doi = {https://doi.org/10.1016/B978-0-12-396500-4.00016-8},
url = {https://www.sciencedirect.com/science/article/pii/B9780123965004000168},
author = {Hugh Griffiths}
}

@article{calib,
doi = {10.3847/1538-4365/aa6df9},
url = {https://dx.doi.org/10.3847/1538-4365/aa6df9},
year = {2017},
publisher = {The American Astronomical Society},
volume = {230},
number = {1},
pages = {7},
author = {R. A. Perley and B. J. Butler},
title = {An Accurate Flux Density Scale from 50 MHz to 50 GHz},
journal = {The Astrophysical Journal Supplement Series}
}

@article{numpy,
author = {van der Walt, Stéfan and Colbert, S. and Varoquaux, Gael},
year = {2011},
month = {05},
pages = {22 - 30},
title = {The NumPy Array: A Structure for Efficient Numerical Computation},
volume = {13},
journal = {Computing in Science \& Engineering},
doi = {10.1109/MCSE.2011.37}
}

@article{ska,
author = {Caiazzo, M},
year = {2017},
month = {07},
pages = {},
title = {SKA phase 1 system requirements specification},
volume = {11},
journal = {SKA Organisation},
url = {\url{https://www.skao.int/sites/default/files/documents/d3-SKA-TEL-SKO-0000008-Rev11_SKA1SystemRequirementSpecification_0.pdf}}
}

@ARTICLE{5164979,  author={Lonsdale, Colin J. and Cappallo, Roger J. and Morales, Miguel F. and Briggs, Frank H. and Benkevitch, Leonid and Bowman, Judd D. and Bunton, John D. and Burns, Steven and Corey, Brian E. and deSouza, Ludi and Doeleman, Sheperd S. and Derome, Mark and Deshpande, Avinash and Gopala, Modavanatt Ramakrishna and Greenhill, Lincoln J. and Herne, David Edwin and Hewitt, Jacqueline N. and Kamini, P. A. and Kasper, Justin C. and Kincaid, Barton B. and Kocz, Jonathan and Kowald, Errol and Kratzenberg, Eric and Kumar, Deepak and Lynch, Mervyn J. and Madhavi, S. and Matejek, Michael and Mitchell, Daniel A. and Morgan, Edward and Oberoi, Divya and Ord, Steven and Pathikulangara, Joseph and Prabu, T. and Rogers, Alan E. E. and Roshi, Anish and Salah, Joseph E. and Sault, Robert J. and Shankar, N. Udaya and Srivani, K. S. and Stevens, Jamie and Tingay, Steven and Vaccarella, Annino and Waterson, Mark and Wayth, Randall B. and Webster, Rachel L. and Whitney, Alan R. and Williams, Andrew and Williams, Christopher},  journal={Proceedings of the IEEE},   title={The Murchison Widefield Array: Design Overview},   year={2009},  volume={97},  number={8},  pages={1497-1506},  doi={10.1109/JPROC.2009.2017564}}

@ARTICLE{astropy,
    author = {{Astropy Collaboration} and {Price-Whelan}, A.~M. and {Sip{\H{o}}cz}, B.~M. and {G{\"u}nther}, H.~M. and {Lim}, P.~L. and {Crawford}, S.~M. and {Conseil}, S. and {Shupe}, D.~L. and {Craig}, M.~W. and {Dencheva}, N. and {Ginsburg}, A. and {VanderPlas}, J.~T. and {Bradley}, L.~D. and {P{\'e}rez-Su{\'a}rez}, D. and {de Val-Borro}, M. and {Aldcroft}, T.~L. and {Cruz}, K.~L. and {Robitaille}, T.~P. and {Tollerud}, E.~J. and {Ardelean}, C. and {Babej}, T. and {Bach}, Y.~P. and {Bachetti}, M. and {Bakanov}, A.~V. and {Bamford}, S.~P. and {Barentsen}, G. and {Barmby}, P. and {Baumbach}, A. and {Berry}, K.~L. and {Biscani}, F. and {Boquien}, M. and {Bostroem}, K.~A. and {Bouma}, L.~G. and {Brammer}, G.~B. and {Bray}, E.~M. and {Breytenbach}, H. and {Buddelmeijer}, H. and {Burke}, D.~J. and {Calderone}, G. and {Cano Rodr{\'\i}guez}, J.~L. and {Cara}, M. and {Cardoso}, J.~V.~M. and {Cheedella}, S. and {Copin}, Y. and {Corrales}, L. and {Crichton}, D. and {D'Avella}, D. and {Deil}, C. and {Depagne}, {\'E}. and {Dietrich}, J.~P. and {Donath}, A. and {Droettboom}, M. and {Earl}, N. and {Erben}, T. and {Fabbro}, S. and {Ferreira}, L.~A. and {Finethy}, T. and {Fox}, R.~T. and {Garrison}, L.~H. and {Gibbons}, S.~L.~J. and {Goldstein}, D.~A. and {Gommers}, R. and {Greco}, J.~P. and {Greenfield}, P. and {Groener}, A.~M. and {Grollier}, F. and {Hagen}, A. and {Hirst}, P. and {Homeier}, D. and {Horton}, A.~J. and {Hosseinzadeh}, G. and {Hu}, L. and {Hunkeler}, J.~S. and {Ivezi{\'c}}, {\v{Z}}. and {Jain}, A. and {Jenness}, T. and {Kanarek}, G. and {Kendrew}, S. and {Kern}, N.~S. and {Kerzendorf}, W.~E. and {Khvalko}, A. and {King}, J. and {Kirkby}, D. and {Kulkarni}, A.~M. and {Kumar}, A. and {Lee}, A. and {Lenz}, D. and {Littlefair}, S.~P. and {Ma}, Z. and {Macleod}, D.~M. and {Mastropietro}, M. and {McCully}, C. and {Montagnac}, S. and {Morris}, B.~M. and {Mueller}, M. and {Mumford}, S.~J. and {Muna}, D. and {Murphy}, N.~A. and {Nelson}, S. and {Nguyen}, G.~H. and {Ninan}, J.~P. and {N{\"o}the}, M. and {Ogaz}, S. and {Oh}, S. and {Parejko}, J.~K. and {Parley}, N. and {Pascual}, S. and {Patil}, R. and {Patil}, A.~A. and {Plunkett}, A.~L. and {Prochaska}, J.~X. and {Rastogi}, T. and {Reddy Janga}, V. and {Sabater}, J. and {Sakurikar}, P. and {Seifert}, M. and {Sherbert}, L.~E. and {Sherwood-Taylor}, H. and {Shih}, A.~Y. and {Sick}, J. and {Silbiger}, M.~T. and {Singanamalla}, S. and {Singer}, L.~P. and {Sladen}, P.~H. and {Sooley}, K.~A. and {Sornarajah}, S. and {Streicher}, O. and {Teuben}, P. and {Thomas}, S.~W. and {Tremblay}, G.~R. and {Turner}, J.~E.~H. and {Terr{\'o}n}, V. and {van Kerkwijk}, M.~H. and {de la Vega}, A. and {Watkins}, L.~L. and {Weaver}, B.~A. and {Whitmore}, J.~B. and {Woillez}, J. and {Zabalza}, V. and {Astropy Contributors}},
        title = "{The Astropy Project: Building an Open-science Project and Status of the v2.0 Core Package}",
      journal = {\aj},
         year = 2018,
        month = sep,
       volume = {156},
       number = {3},
          eid = {123},
        pages = {123},
          doi = {10.3847/1538-3881/aabc4f},
archivePrefix = {arXiv},
       eprint = {1801.02634},
 primaryClass = {astro-ph.IM},
       adsurl = {https://ui.adsabs.harvard.edu/abs/2018AJ....156..123A}
}

@ARTICLE{mwa,
       author = {{Tingay}, S.~J. and {Goeke}, R. and {Bowman}, J.~D. and {Emrich}, D. and {Ord}, S.~M. and {Mitchell}, D.~A. and {Morales}, M.~F. and {Booler}, T. and {Crosse}, B. and {Wayth}, R.~B. and {Lonsdale}, C.~J. and {Tremblay}, S. and {Pallot}, D. and {Colegate}, T. and {Wicenec}, A. and {Kudryavtseva}, N. and {Arcus}, W. and {Barnes}, D. and {Bernardi}, G. and {Briggs}, F. and {Burns}, S. and {Bunton}, J.~D. and {Cappallo}, R.~J. and {Corey}, B.~E. and {Deshpande}, A. and {Desouza}, L. and {Gaensler}, B.~M. and {Greenhill}, L.~J. and {Hall}, P.~J. and {Hazelton}, B.~J. and {Herne}, D. and {Hewitt}, J.~N. and {Johnston-Hollitt}, M. and {Kaplan}, D.~L. and {Kasper}, J.~C. and {Kincaid}, B.~B. and {Koenig}, R. and {Kratzenberg}, E. and {Lynch}, M.~J. and {Mckinley}, B. and {Mcwhirter}, S.~R. and {Morgan}, E. and {Oberoi}, D. and {Pathikulangara}, J. and {Prabu}, T. and {Remillard}, R.~A. and {Rogers}, A.~E.~E. and {Roshi}, A. and {Salah}, J.~E. and {Sault}, R.~J. and {Udaya-Shankar}, N. and {Schlagenhaufer}, F. and {Srivani}, K.~S. and {Stevens}, J. and {Subrahmanyan}, R. and {Waterson}, M. and {Webster}, R.~L. and {Whitney}, A.~R. and {Williams}, A. and {Williams}, C.~L. and {Wyithe}, J.~S.~B.},
        title = "{The Murchison Widefield Array: The Square Kilometre Array Precursor at Low Radio Frequencies}",
      journal = {\pasa},
         year = 2013,
        month = jan,
       volume = {30},
          eid = {e007},
        pages = {e007},
          doi = {10.1017/pasa.2012.007},
       adsurl = {https://ui.adsabs.harvard.edu/abs/2013PASA...30....7T}
}

@ARTICLE{askap,
       author = {{Johnston}, S. and {Taylor}, R. and {Bailes}, M. and {Bartel}, N. and {Baugh}, C. and {Bietenholz}, M. and {Blake}, C. and {Braun}, R. and {Brown}, J. and {Chatterjee}, S. and {Darling}, J. and {Deller}, A. and {Dodson}, R. and {Edwards}, P. and {Ekers}, R. and {Ellingsen}, S. and {Feain}, I. and {Gaensler}, B. and {Haverkorn}, M. and {Hobbs}, G. and {Hopkins}, A. and {Jackson}, C. and {James}, C. and {Joncas}, G. and {Kaspi}, V. and {Kilborn}, V. and {Koribalski}, B. and {Kothes}, R. and {Landecker}, T. and {Lenc}, E. and {Lovell}, J. and {Macquart}, J. -P. and {Manchester}, R. and {Matthews}, D. and {McClure-Griffiths}, N. and {Norris}, R. and {Pen}, U. -L. and {Phillips}, C. and {Power}, C. and {Protheroe}, R. and {Sadler}, E. and {Schmidt}, B. and {Stairs}, I. and {Staveley-Smith}, L. and {Stil}, J. and {Tingay}, S. and {Tzioumis}, A. and {Walker}, M. and {Wall}, J. and {Wolleben}, M.},
        title = "{Science with ASKAP. The Australian square-kilometre-array pathfinder}",
      journal = {Experimental Astronomy},
         year = 2008,
        month = dec,
       volume = {22},
       number = {3},
        pages = {151-273},
          doi = {10.1007/s10686-008-9124-7},
       adsurl = {https://ui.adsabs.harvard.edu/abs/2008ExA....22..151J}
}

@ARTICLE{meerkat,
       author = {{Jonas}, J.~L.},
        title = "{MeerKAT - The South African Array With Composite Dishes and Wide-Band Single Pixel Feeds}",
      journal = {IEEE Proceedings},
         year = 2009,
        month = aug,
       volume = {97},
       number = {8},
        pages = {1522-1530},
          doi = {10.1109/JPROC.2009.2020713},
       adsurl = {https://ui.adsabs.harvard.edu/abs/2009IEEEP..97.1522J}
}

@ARTICLE{vla,
       author = {{Condon}, J.~J. and {Cotton}, W.~D. and {Greisen}, E.~W. and {Yin}, Q.~F. and {Perley}, R.~A. and {Taylor}, G.~B. and {Broderick}, J.~J.},
        title = "{The NRAO VLA Sky Survey}",
      journal = {\aj},
         year = 1998,
        month = may,
       volume = {115},
       number = {5},
        pages = {1693-1716},
          doi = {10.1086/300337},
       adsurl = {https://ui.adsabs.harvard.edu/abs/1998AJ....115.1693C}
}

@article{starlink_bassa,
	author = {{Bassa, C. G.} and {Di Vruno, F.} and {Winkel, B.} and {Józsa, G. I. G.} and {Brentjens, M. A.} and {Zhang, X.}},
	title = {Bright unintended electromagnetic radiation from second-generation Starlink satellites},
	DOI= "10.1051/0004-6361/202451856",
	url= "https://doi.org/10.1051/0004-6361/202451856",
	journal = {A\&A},
	year = 2024,
	volume = 689,
	pages = "L10",
}

@ARTICLE{eda2_old_calibration,
       author = {{Benthem}, P. and {Wayth}, R. and {de Lera Acedo}, E. and {Zarb Adami}, K. and {Alderighi}, M. and {Belli}, C. and {Bolli}, P. and {Booler}, T. and {Borg}, J. and {Broderick}, J.~W. and {Chiarucci}, S. and {Chiello}, R. and {Ciani}, L. and {Comoretto}, G. and {Crosse}, B. and {Davidson}, D. and {DeMarco}, A. and {Emrich}, D. and {van Es}, A. and {Fierro}, D. and {Faulkner}, A. and {Gerbers}, M. and {Razavi-Ghods}, N. and {Hall}, P. and {Horsley}, L. and {Juswardy}, B. and {Kenney}, D. and {Steele}, K. and {Magro}, A. and {Mattana}, A. and {McKinley}, B. and {Monari}, J. and {Naldi}, G. and {Nanni}, J. and {Di Ninni}, P. and {Paonessa}, F. and {Perini}, F. and {Poloni}, M. and {Pupillo}, G. and {Rusticelli}, S. and {Schiaffino}, M. and {Schillir{\`o}}, F. and {Schnetler}, H. and {Singuaroli}, R. and {Sokolowski}, M. and {Sutinjo}, A. and {Tartarini}, G. and {Ung}, D. and {Bij de Vaate}, J.~G. and {Virone}, G. and {Walker}, M. and {Waterson}, M. and {Wijnholds}, S.~J. and {Williams}, A.},
        title = "{The Aperture Array Verification System 1: System overview and early commissioning results}",
      journal = {\aap},
         year = 2021,
        month = nov,
       volume = {655},
          eid = {A5},
        pages = {A5},
          doi = {10.1051/0004-6361/202040086},
archivePrefix = {arXiv},
       eprint = {2110.03217},
 primaryClass = {astro-ph.IM},
       adsurl = {https://ui.adsabs.harvard.edu/abs/2021A&A...655A...5B}
}

\newpage

\appendix

\section{Sinusoidal motion in measured positions}
\label{app:sinusoidal}

\begin{figure*}[hbt!]
\centering
\includegraphics[width=1\linewidth]{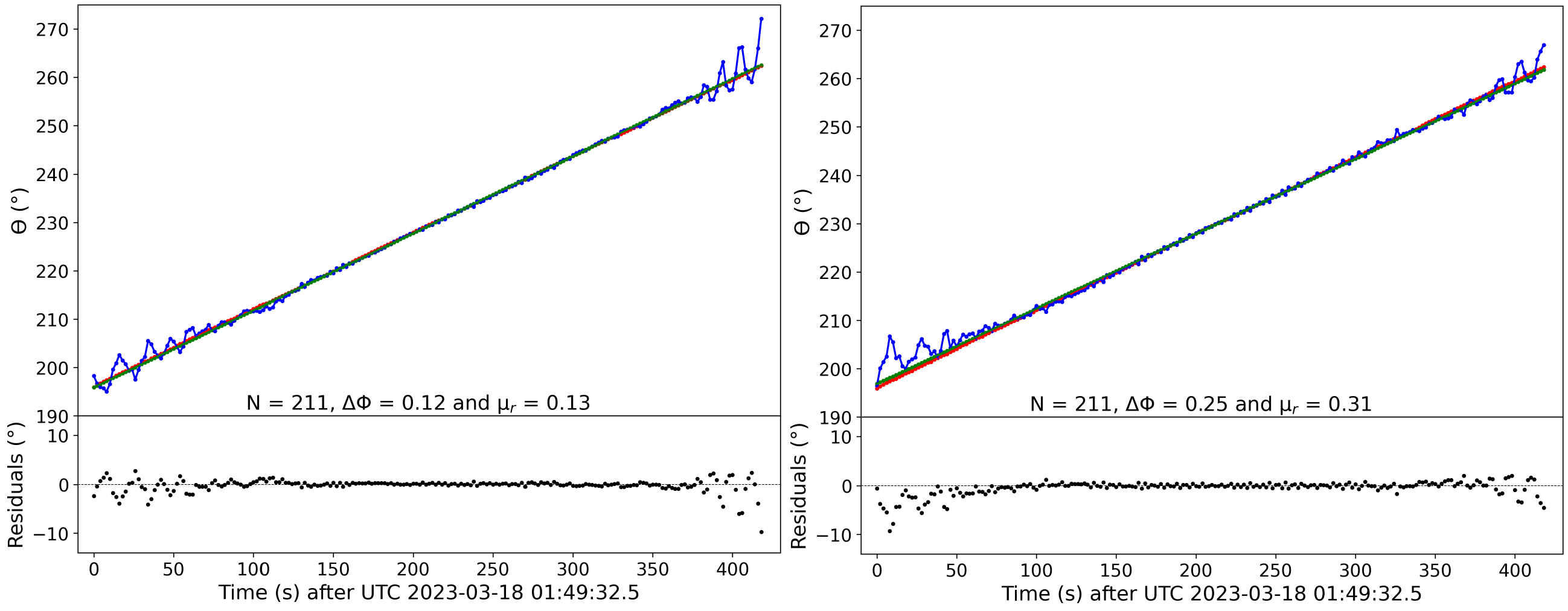}
\caption{Two scatter plots showing a real example of detections of ORBCOMM FM113 in the 137.5 MHz dataset (left) and a simulated pass of ORBCOMM FM113 (right). The scatter plots show $\theta_{P}$ (red points) and $\theta_{M}$ (blue points). The green points are from the linear line of best fit for the $\theta_{M}$ values. The residuals are between the $\theta_{M}$ and $\theta_{P}$ values. The image number is zeroed on the image of the first candidate.}
\label{fig:sinusoidal_motion}
\end{figure*}

This Appendix investigates a phenomenon, where the offset between the predicted and measured locations of a satellite varied sinusoidally in time for some cases. This effect was strongest at low elevations. An example of this can be seen in the left panel in Figure \ref{fig:sinusoidal_motion}, as it also manifests in the values of $\theta_{P}$ and $\theta_{M}$.

There are two effects here which are important to separate. The first is due to photometric centroid shift. Image three in Figure \ref{fig:dif_images} shows an ideal separation between the positive and negative intensities of the satellite due to time differencing. When the separation between these two decreases, it can cause an asymmetry in the intensity profile of the positive signal, which results in a shift in its centroid. The closer these two are, the greater this observed shift is. 

The second effect is due to sub-pixel sampling of the fitted centroid. At frequencies of 73.4 MHz and 325 MHz, the sampling of pixels is 2.2 and 0.5 degrees/pixel respectively. At lower frequencies, it can take a satellite several images to fully cross a pixel. This sub-pixel sampling imparts a sinusoidal pattern on the offsets between the predicted and measured positions over time.

To test these effects, a simulation was run. A pass of the satellite ORBCOMM FM113 (NORAD 41185) was chosen, as its plotted values for $\theta_{M}$ showed this sinusoidal variation as can be seen in the left panel in Figure \ref{fig:sinusoidal_motion}. To recreate this as simply as possible, the images were replaced with arrays of zeros, and a synthetic signal was introduced at the predicted locations of the satellite. This included a positive signal at the location in the current image, and a negative signal at the location of the satellite in the subtracted image used in the time differencing. The injected signal was a circular Gaussian multiplied with a Bessel function as an approximation of the point spread function. This was parameterised as follows:

\begin{equation}\label{eqn:bessel1}
    I_{\text{model}} = J_0\left(\frac{r}{\sigma_{\text{B}}}\right) \cdot A \exp\left(-\frac{1}{2 \sigma_{\text{G}}^{2}} \left[(x - x_{0})^{2} + (y - y_{0})^{2}\right]\right),
\end{equation}

where $r = \sqrt{(x - x_0)^2 + (y - y_0)^2}$ is the radial distance from the centre $(x_0, y_0)$, $A$ is the amplitude, $\sigma_{\text{B}}$ is the standard deviation of the Bessel function, $\sigma_{\text{G}}$ is the standard deviation of the Gaussian function and $J_0(x)$ is a zero-order Bessel function of the first kind:

\begin{equation}
    J_0(a) = \sum_{n=0}^{\infty} \frac{(-1)^n a^{2n}}{2^{2n}(n!)^2}.
\end{equation}

The following values estimated from the real measurements of the pass of ORBCOMM FM113: $A = 50,000,000$, $(x_0, y_0)$ were the TLE predicted positions of the satellite, $\sigma_{\text{B}}$ was derived from the size of the PSF in degrees for this frequency, and $\sigma_{\text{G}} = 4$ (appropriate taper for this frequency).

The right panel in Figure \ref{fig:sinusoidal_motion} shows that the simulation displays the same sinusoidal pattern at the beginning and end of the pass of the satellite as the real case on the left. This effect can therefore be explained with geometry and is not related to fluctuations in the position of the received signal.

The photometric centroid shift is minimised by choosing a large enough time offset between the two images being used in the time difference, and imposing an elevation cutoff helps when the satellite is moving slower over the sky at low elevations. The sub-pixel sampling effect is minimised by fitting the straight line (green line in Figure \ref{fig:sinusoidal_motion}) to the $\theta_{P}$ and $\theta_{M}$ values. This performs well in real examples.

In summary, understanding both the photometric centroid shift and sub-pixel sampling effects allows us to manage their effects to ensure that satellite identification remains robust and accurate.

\section{Technical details for making identifications.}
\label{app:identifications}

This Appendix provides the quantitative, technical analysis of how identifications are made to accompany Section \ref{making_identifications}.

To repeat for context, we now refer to the TLE predicted position of a satellite as the `predicted' position, and the position of the fitted Gaussian ($x_{0}$ and $y_{0}$) as the `measured' position. Each candidate will have this predicted and measured position information.

\begin{figure}[hbt!]
\centering
\includegraphics[width=1\linewidth]{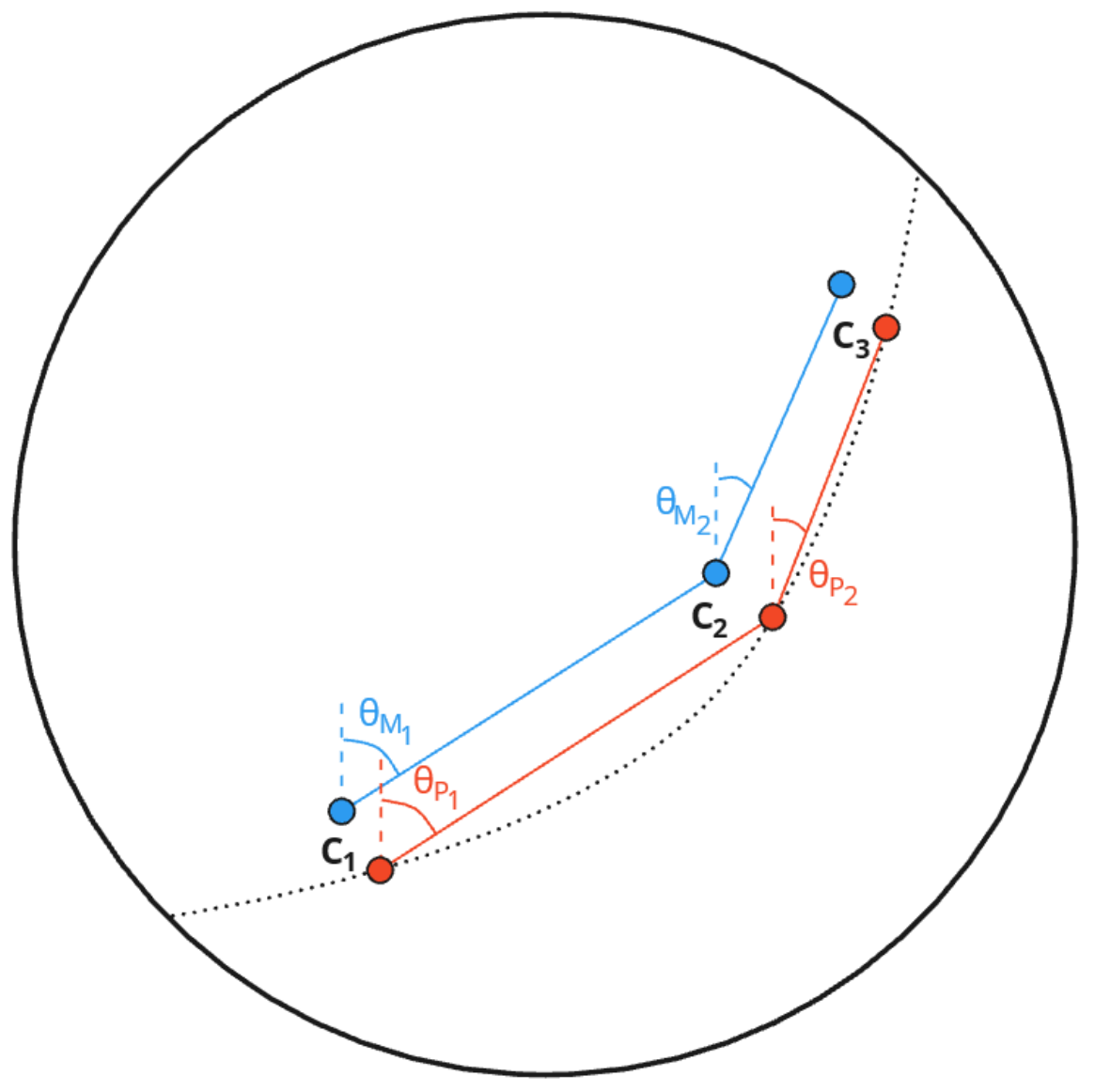}
\caption{An illustration showing how $\theta_{M}$ and $\theta_{P}$ are calculated for three candidate detections $C_{1}$, $C_{2}$ and $C_{3}$ from three separate images. The circular outline shows the full visible sky horizon to horizon and the curved black dotted line is the predicted path of a satellite across the sky from TLE information. These candidates are for the same satellite, where the blue points mark the measured locations, and the red points mark the predicted locations. The distance between these is exaggerated in this figure.}
\label{fig:sat_diagram_trig}
\end{figure}

The instantaneous direction of travel of a satellite over the sky in an image is calculated as the bearing angle between two adjacent candidates and is denoted by $\theta$. $\theta$ is the bearing angle measured clockwise between $\hat{y}$, a unit vector in the y direction, and $\overrightarrow{C_{1}C_{2}}$, the vector between the position of the candidate at time $t_{1}$ ($C_{1}$) and the position of the next candidate at time $t_{2}$ ($C_{2}$). $\theta$ is calculated for both the predicted ($\theta_{P}$) and measured ($\theta_{M}$) positions for each candidate, resulting in $N-1$ $\theta$s for $N$ candidates. This process is illustrated in Figure \ref{fig:sat_diagram_trig}.

The list of candidates is then sorted so that all candidates for each satellite are put into 15 minute bins which are designed to be long enough that any object in the TLE list will pass over the sky completely during this time, and where each bin represents individual passes over the sky for each satellite. The number of candidates for the same satellite in each bin were summed ($N$) and stored with each candidate. If $N < 4$, then these candidates would be discarded, as the algorithm performs best when $N$ is high. There was no evidence that certain satellites transmitted less frequently than this cadence over the course of a pass\footnote{The only caveat to this was Starlink satellites at 137.5 MHz which have been found to transmit every 100 s in \citep{grigg_starlink}. As these have already been reported on in the literature, we omit these from the analyses in this paper for the 137.5 MHz dataset.}.

\begin{figure*}[hbt!]
\centering
\includegraphics[width=1\linewidth]{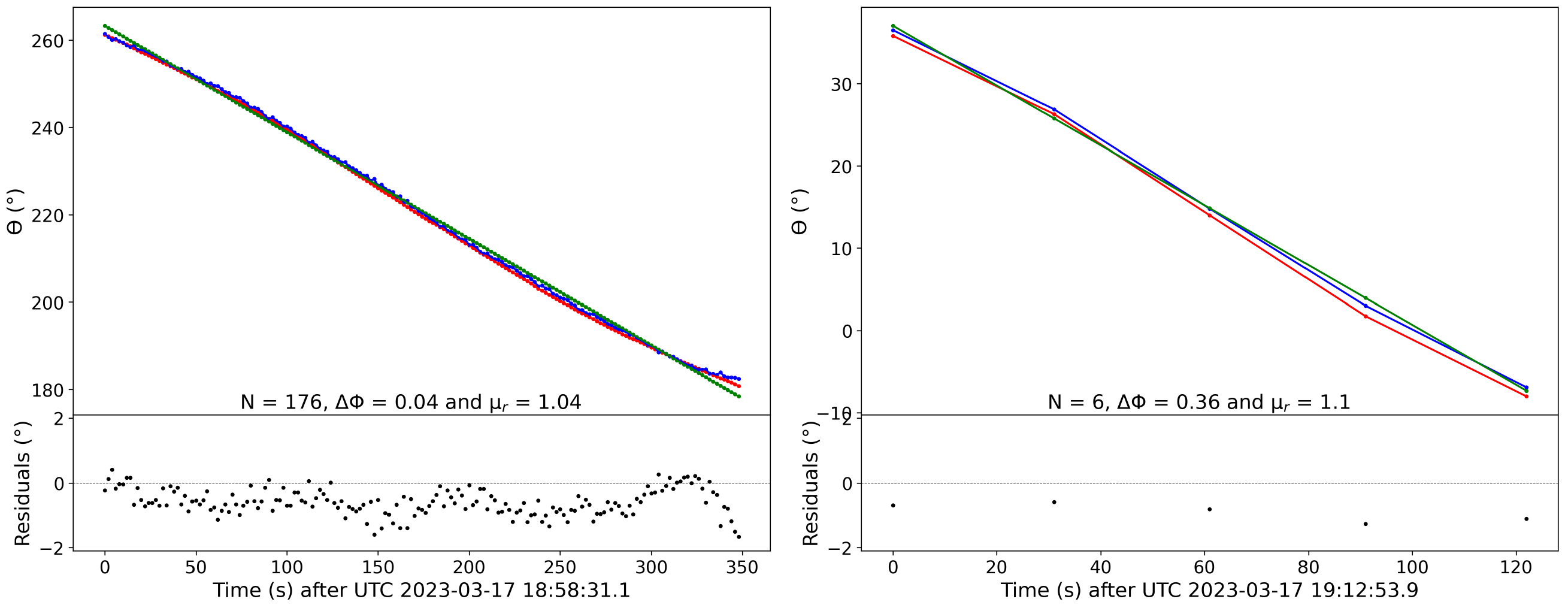}
\caption{A plot showing $\theta_{P}$ (red points) and $\theta_{M}$ (blue points) calculated for passes of ORBCOMM FM117 (left) and SPACEBEE-111 (right). The green points are from the linear line of best fit for the $\theta_{M}$ values. The residuals are between $\theta_{M}$ and $\theta_{P}$ values. The image number is zeroed on the first image the candidate is detected. $\phi$ is measured as the clockwise bearing angle of each of the lines of best fit.}
\label{fig:theta_graph}
\end{figure*}

$\theta_{P}$ and $\theta_{M}$ are then calculated for each group of candidates. Figure \ref{fig:theta_graph} shows real examples of what this looks like for the pass of a satellite that transmits every consecutive image across its path (left) and a satellite that transmits periodically across its path (right). $\theta_{P}$ and $\theta_{M}$ gradually increase or decrease depending on the direction of travel of the satellite over the sky. 

There is uncertainty in the measured positions of the satellite (due to noise and fitting) and uncertainty in the TLE predicted positions (described in detail for LEO satellites in \citet{tle_uncert}). Although these will perturb the $\theta_{P}$ and $\theta_{M}$ values in Figure \ref{fig:theta_graph}, they will still both follow the same overall direction for a correct identification. Fitting a straight line to $\theta_{P}$ and $\theta_{M}$ provides a way to `smooth' out these uncertainties, and proves to be robust for low $N$. This being said, the algorithm performs best when $N$ is high.

The bearing angle measured clockwise from a vertical unit vector in the $\theta$ - time plane (shown in Figure \ref{fig:theta_graph}) with the fitted line across all $\theta_{P}$ and $\theta_{M}$ is calculated and referred to as $\phi_{P}$ and $\phi_{M}$, respectively. The difference in these values, $|\phi_{P} - \phi_{M}|$, or $\Delta\phi$, shows the variation in travel direction over the sky between the measured and predicted positions. This value is stored with every candidate detection.

The next step was to reduce the number of candidates by ensuring that no candidates had the same location and time. The signals detected in the data cover at least the width of the PSF on the sky, or even longer if in LEO. This means that there are likely going to be multiple satellites over this PSF width, for which multiple may have satisfied the candidate criteria so far. Figure \ref{fig:sat_diagram_path_comparison} shows a simplified example of this.

Each candidate now has $N$ and $\Delta\phi$ calculated. If there was more than one candidate for the same time and location, only the candidate with the highest $N$ was kept. If $N$ was equal for two candidates, then the candidate with the lowest value for $\Delta\phi$ was kept. Therefore, in Figure \ref{fig:sat_diagram_path_comparison}, the six candidates for satellite `x' would be discarded, because their $N$ (six) is less than the $N$ (twenty) of satellite `y'.

Once these discards had been made for the whole dataset, each candidate had $N$ and $\Delta\phi$ recalculated within each 15 minute bin. Candidates with $N < 4$ were again discarded, as well as candidates with $\Delta\phi > 10$.

The mean of the residuals between the fitted line to $\theta_{M}$ and the $\theta_{P}$ values was calculated and denoted as $\mu_{r}$. When the predicted trajectories of two satellites are very similar in position, but offset in time, this value of $\mu_{r}$ will become larger. In Figure \ref{fig:theta_graph}, the fitted values for $\theta_{P}$ and $\theta_{M}$ will be offset by a fixed distance proportional to this timing offset. Candidates with $\mu_{r} > 3^{\circ}$ are therefore discarded. This value was data driven and derived by observing examples where two satellites had very similar orbits. These identifications are then stored in the database.

An important caveat is that this algorithm was parameterised to prioritise making higher confidence identifications. Some of the parameters could have been relaxed, which would have made more identifications, especially in the noisier datasets, but at the expense of allowing more misidentifications into the results.

\section{Satellite detection list by NORAD}
\label{app:sat_table}

This Appendix shows in detail the list of identifications in full in Table \ref{tab:detections_by_norad}.

\newpage
\begin{table*}[htbp]
\caption{Individual identifications per NORAD I.D. for each dataset}
\label{tab:detections_by_norad}
\begin{tabular}{ccccccccc}

\hline
\hline
Frequency &
  \begin{tabular}[c]{@{}c@{}}Dataset\\ start date\end{tabular} &
  Telescope &
  \begin{tabular}[c]{@{}c@{}}NORAD\\ I.D.\end{tabular} &
  Name &
  \begin{tabular}[c]{@{}c@{}}Identifications\\ XX\end{tabular} &
  \begin{tabular}[c]{@{}c@{}}Identifications\\ YY\end{tabular} &
  \begin{tabular}[c]{@{}c@{}}Closest\\ range (km)\end{tabular} &
  \begin{tabular}[c]{@{}c@{}}Furthest\\ range (km)\end{tabular} \\
\hline
\hline

96.9                         & 2022-12-18 18:22:15.5 & EDA2                        & 25544 & ISS            & 68  & 72  & 439  & 665  \\
96.9                         & 2022-12-16 10:10:30.5 & EDA2                        & 25544 & ISS            & 66  & 64  & 447  & 593  \\
\hline
98.4                         & 2022-10-18 18:25:12.5 & EDA2                        & 25544 & ISS            & 189 & 170 & 422  & 978  \\
\hline
136.7                        & 2023-04-25 01:04:37.5 & EDA2                        & 1291  & SOLRAD 7B      & 470 & 469 & 934  & 2015 \\
136.7                        & 2023-04-25 01:04:37.5 & EDA2                        & 41182 & ORBCOMM FM 110 & 290 & 292 & 866  & 1593 \\
\hline
137.5                        & 2023-03-17 12:19:09.0 & EDA2                        & 40087 & ORBCOMM FM 107 & 566 & 563 & 722  & 1682 \\
137.5                        & 2023-03-17 12:19:09.0 & EDA2                        & 41187 & ORBCOMM FM 108 & 557 & 559 & 716  & 1649 \\
137.5                        & 2023-03-17 12:19:09.0 & EDA2                        & 40086 & ORBCOMM FM 109 & 556 & 559 & 757  & 1668 \\
137.5                        & 2023-03-17 12:19:09.0 & EDA2                        & 41188 & ORBCOMM FM 117 & 556 & 557 & 737  & 1658 \\
137.5                        & 2023-03-17 12:19:09.0 & EDA2                        & 41179 & ORBCOMM FM 114 & 543 & 544 & 906  & 1654 \\
137.5                        & 2023-03-17 12:19:09.0 & EDA2                        & 33591 & NOAA 19        & 473 & 476 & 922  & 1991 \\
137.5                        & 2023-03-17 12:19:09.0 & EDA2                        & 41183 & ORBCOMM FM 118 & 453 & 443 & 719  & 1634 \\
137.5                        & 2023-03-17 12:19:09.0 & EDA2                        & 28654 & NOAA 18        & 444 & 444 & 965  & 1871 \\
137.5                        & 2023-03-17 12:19:09.0 & EDA2                        & 41182 & ORBCOMM FM 110 & 398 & 389 & 722  & 1664 \\
137.5                        & 2023-03-17 12:19:09.0 & EDA2                        & 41189 & ORBCOMM FM 116 & 305 & 304 & 794  & 1638 \\
137.5                        & 2023-03-17 12:19:09.0 & EDA2                        & 25338 & NOAA 15        & 264 & 268 & 1229 & 1828 \\
137.5                        & 2023-03-17 12:19:09.0 & EDA2                        & 4237  & OPS 7613       & 231 & 251 & 1043 & 1976 \\
137.5                        & 2023-03-17 12:19:09.0 & EDA2                        & 41185 & ORBCOMM FM 113 & 210 & 210 & 746  & 1593 \\
137.5                        & 2023-03-17 12:19:09.0 & EDA2                        & 41184 & ORBCOMM FM 112 & 175 & 177 & 1069 & 1686 \\
137.5                        & 2023-03-17 12:19:09.0 & EDA2                        & 5680  & OPS 7898       & 106 & 96  & 1076 & 1490 \\
137.5                        & 2023-03-17 12:19:09.0 & EDA2                        & 48897 & SPACEBEE-109   & 92  & 93  & 559  & 1173 \\
137.5                        & 2023-03-17 12:19:09.0 & EDA2                        & 52179 & SPACEBEE-129   & 64  & 55  & 464  & 1084 \\
137.5                        & 2023-03-17 12:19:09.0 & EDA2                        & 55101 & SPACEBEE-159   & 55  & 61  & 603  & 1224 \\
137.5                        & 2023-03-17 12:19:09.0 & EDA2                        & 48893 & SPACEBEE-106   & 47  & 50  & 901  & 1216 \\
137.5                        & 2023-03-17 12:19:09.0 & EDA2                        & 55087 & SPACEBEE-167   & 45  & 41  & 684  & 1251 \\
137.5                        & 2023-03-17 12:19:09.0 & EDA2                        & 47459 & SPACEBEE-55    & 42  & 43  & 511  & 1251 \\
137.5                        & 2023-03-17 12:19:09.0 & EDA2                        & 25418 & ORBCOMM FM 15  & 41  & 39  & 1278 & 1700 \\
137.5                        & 2023-03-17 12:19:09.0 & EDA2                        & 48886 & SPACEBEE-102   & 40  & 39  & 799  & 1195 \\
137.5                        & 2023-03-17 12:19:09.0 & EDA2                        & 52187 & SPACEBEE-135   & 36  & 38  & 467  & 1130 \\
137.5                        & 2023-03-17 12:19:09.0 & EDA2                        & 52185 & SPACEBEE-132   & 32  & 31  & 484  & 1126 \\
137.5                        & 2023-03-17 12:19:09.0 & EDA2                        & 48896 & SPACEBEE-105   & 30  & 31  & 503  & 1113 \\
137.5                        & 2023-03-17 12:19:09.0 & EDA2                        & 52182 & SPACEBEE-128   & 29  & 30  & 565  & 1140 \\
137.5                        & 2023-03-17 12:19:09.0 & EDA2                        & 55095 & SPACEBEE-162   & 29  & 26  & 549  & 1206 \\
137.5                        & 2023-03-17 12:19:09.0 & EDA2                        & 48890 & SPACEBEE-104   & 27  & 27  & 606  & 913  \\
137.5                        & 2023-03-17 12:19:09.0 & EDA2                        & 52394 & SPACEBEE-143   & 27  & 26  & 590  & 1208 \\
137.5                        & 2023-03-17 12:19:09.0 & EDA2                        & 52010 & SPACEBEE-114   & 27  & 24  & 592  & 1111 \\
137.5                        & 2023-03-17 12:19:09.0 & EDA2                        & 52012 & SPACEBEE-118   & 27  & 24  & 587  & 1217 \\
137.5                        & 2023-03-17 12:19:09.0 & EDA2                        & 52399 & SPACEBEE-140   & 26  & 26  & 515  & 1165 \\
137.5                        & 2023-03-17 12:19:09.0 & EDA2                        & 52180 & SPACEBEE-134   & 26  & 26  & 471  & 1085 \\
137.5                        & 2023-03-17 12:19:09.0 & EDA2                        & 52009 & SPACEBEE-121   & 26  & 24  & 655  & 1167 \\
137.5                        & 2023-03-17 12:19:09.0 & EDA2                        & 52411 & SPACEBEE-145   & 25  & 18  & 586  & 1162 \\
137.5                        & 2023-03-17 12:19:09.0 & EDA2                        & 52409 & SPACEBEE-150   & 23  & 21  & 607  & 1160 \\
137.5                        & 2023-03-17 12:19:09.0 & EDA2                        & 52416 & SPACEBEE-155   & 24  & 26  & 750  & 1161 \\
137.5                        & 2023-03-17 12:19:09.0 & EDA2                        & 52401 & SPACEBEE-142   & 22  & 24  & 662  & 1199 \\
137.5                        & 2023-03-17 12:19:09.0 & EDA2                        & 22491 & PEGASUS R/B    & 22  & 22  & 1495 & 1563 \\
137.5                        & 2023-03-17 12:19:09.0 & EDA2                        & 52400 & SPACEBEE-141   & 22  & 22  & 975  & 1086 \\
137.5                        & 2023-03-17 12:19:09.0 & EDA2                        & 48884 & SPACEBEE-101   & 22  & 21  & 663  & 1021 \\
137.5                        & 2023-03-17 12:19:09.0 & EDA2                        & 52015 & SPACEBEE-119   & 19  & 20  & 821  & 1177 \\
137.5                        & 2023-03-17 12:19:09.0 & EDA2                        & 52415 & SPACEBEE-147   & 19  & 18  & 765  & 1203 \\
137.5                        & 2023-03-17 12:19:09.0 & EDA2                        & 52025 & SPACEBEE-112   & 18  & 23  & 654  & 955  \\
\hline
\end{tabular}
\end{table*}

\clearpage
\begin{table*}[htbp]
\ContinuedFloat
\caption{Individual identifications per NORAD I.D. for each dataset (continued)}
\begin{tabular}{ccccccccc}

\hline
\hline
Frequency &
  \begin{tabular}[c]{@{}c@{}}Dataset\\ start date\end{tabular} &
  Telescope &
  \begin{tabular}[c]{@{}c@{}}NORAD\\ I.D.\end{tabular} &
  Name &
  \begin{tabular}[c]{@{}c@{}}Identifications\\ XX\end{tabular} &
  \begin{tabular}[c]{@{}c@{}}Identifications\\ YY\end{tabular} &
  \begin{tabular}[c]{@{}c@{}}Closest\\ range (km)\end{tabular} &
  \begin{tabular}[c]{@{}c@{}}Furthest\\ range (km)\end{tabular} \\
\hline
\hline

137.5                        & 2023-03-17 12:19:09.0 & EDA2                        & 52181 & SPACEBEE-131   & 18  & 18  & 807  & 977  \\
137.5                        & 2023-03-17 12:19:09.0 & EDA2                        & 52014 & SPACEBEE-122   & 16  & 16  & 1044 & 1156 \\
137.5                        & 2023-03-17 12:19:09.0 & EDA2                        & 52023 & SPACEBEE-116   & 15  & 15  & 737  & 1014 \\
137.5                        & 2023-03-17 12:19:09.0 & EDA2                        & 52029 & SPACEBEE-124   & 14  & 24  & 927  & 1077 \\
137.5                        & 2023-03-17 12:19:09.0 & EDA2                        & 25476 & ORBCOMM FM 22  & 14  & 13  & 811  & 914  \\
137.5                        & 2023-03-17 12:19:09.0 & EDA2                        & 52186 & SPACEBEE-137   & 13  & 12  & 532  & 1140 \\
137.5                        & 2023-03-17 12:19:09.0 & EDA2                        & 52412 & SPACEBEE-146   & 13  & 8   & 721  & 1212 \\
137.5                        & 2023-03-17 12:19:09.0 & EDA2                        & 48895 & SPACEBEE-108   & 12  & 21  & 810  & 1179 \\
137.5                        & 2023-03-17 12:19:09.0 & EDA2                        & 48894 & SPACEBEE-107   & 12  & 12  & 1028 & 1164 \\
137.5                        & 2023-03-17 12:19:09.0 & EDA2                        & 52164 & SPACEBEE-136   & 12  & 9   & 675  & 916  \\
137.5                        & 2023-03-17 12:19:09.0 & EDA2                        & 55099 & SPACEBEE-156   & 11  & 8   & 551  & 1231 \\
137.5                        & 2023-03-17 12:19:09.0 & EDA2                        & 52414 & SPACEBEE-152   & 10  & 16  & 919  & 1142 \\
137.5                        & 2023-03-17 12:19:09.0 & EDA2                        & 52417 & SPACEBEE-154   & 10  & 10  & 631  & 879  \\
137.5                        & 2023-03-17 12:19:09.0 & EDA2                        & 25118 & ORBCOMM FM 6   & 10  & 8   & 1630 & 1713 \\
137.5                        & 2023-03-17 12:19:09.0 & EDA2                        & 25544 & ISS            & 10  & 8   & 448  & 536  \\
137.5                        & 2023-03-17 12:19:09.0 & EDA2                        & 52407 & SPACEBEE-149   & 10  & 6   & 542  & 945  \\
137.5                        & 2023-03-17 12:19:09.0 & EDA2                        & 48888 & SPACEBEE-103   & 9   & 16  & 532  & 1128 \\
137.5                        & 2023-03-17 12:19:09.0 & EDA2                        & 47442 & SPACEBEE-63    & 8   & 11  & 526  & 1146 \\
137.5                        & 2023-03-17 12:19:09.0 & EDA2                        & 52408 & SPACEBEE-153   & 8   & 7   & 754  & 1196 \\
137.5                        & 2023-03-17 12:19:09.0 & EDA2                        & 52176 & SPACEBEE-130   & 7   & 7   & 492  & 1032 \\
137.5                        & 2023-03-17 12:19:09.0 & EDA2                        & 55097 & SPACEBEE-160   & 7   & 0   & 518  & 1076 \\
137.5                        & 2023-03-17 12:19:09.0 & EDA2                        & 52022 & SPACEBEE-117   & 6   & 7   & 680  & 1121 \\
137.5                        & 2023-03-17 12:19:09.0 & EDA2                        & 48904 & SPACEBEE-111   & 6   & 6   & 651  & 915  \\
137.5                        & 2023-03-17 12:19:09.0 & EDA2                        & 52166 & SPACEBEE-138   & 6   & 6   & 583  & 1022 \\
137.5                        & 2023-03-17 12:19:09.0 & EDA2                        & 52011 & SPACEBEE-126   & 5   & 5   & 732  & 874  \\
137.5                        & 2023-03-17 12:19:09.0 & EDA2                        & 52177 & SPACEBEE-139   & 5   & 0   & 734  & 863  \\
137.5                        & 2023-03-17 12:19:09.0 & EDA2                        & 52016 & SPACEBEE-115   & 0   & 51  & 528  & 1238 \\
137.5                        & 2023-03-17 12:19:09.0 & EDA2                        & 25477 & ORBCOMM FM 23  & 0   & 14  & 890  & 982  \\
137.5                        & 2023-03-17 12:19:09.0 & EDA2                        & 48883 & SPACEBEE-100   & 0   & 7   & 831  & 1085 \\
137.5                        & 2023-03-17 12:19:09.0 & EDA2                        & 52413 & SPACEBEE-151   & 0   & 5   & 511  & 724  \\
\hline
146.1                        & 2021-11-18 01:08:33.0 & EDA2                        & 14781 & OSCAR 11       & 283 & 233 & 656  & 1483 \\
146.1                        & 2021-11-18 01:08:33.0 & EDA2                        & 32789 & DELFI C3       & 266 & 250 & 564  & 1318 \\
146.1                        & 2021-11-18 01:08:33.0 & EDA2                        & 40909 & XW-2E          & 259 & 252 & 552  & 1289 \\
146.1                        & 2021-11-18 01:08:33.0 & EDA2                        & 40911 & XW-2B          & 219 & 241 & 562  & 1303 \\
146.1                        & 2021-11-18 01:08:33.0 & EDA2                        & 42778 & MAX VALIER SAT & 218 & 222 & 496  & 1213 \\
146.1                        & 2021-11-18 01:08:33.0 & EDA2                        & 42761 & ZHUHAI-1 01    & 216 & 229 & 838  & 1301 \\
146.1                        & 2021-11-18 01:08:33.0 & EDA2                        & 40903 & XW-2A          & 208 & 198 & 434  & 1098 \\
146.1                        & 2021-11-18 01:08:33.0 & EDA2                        & 43017 & AO-91          & 198 & 190 & 1374 & 1821 \\
146.1                        & 2021-11-18 01:08:33.0 & EDA2                        & 42017 & NAYIF 1        & 193 & 203 & 750  & 1206 \\
146.1                        & 2021-11-18 01:08:33.0 & EDA2                        & 40906 & XW-2C          & 193 & 195 & 642  & 1299 \\
146.1                        & 2021-11-18 01:08:33.0 & EDA2                        & 40907 & XW-2D          & 178 & 175 & 560  & 1284 \\
146.1                        & 2021-11-18 01:08:33.0 & EDA2                        & 41999 & BGUSAT         & 175 & 160 & 620  & 1185 \\
146.1                        & 2021-11-18 01:08:33.0 & EDA2                        & 40910 & XW-2F          & 169 & 159 & 547  & 1266 \\
146.1                        & 2021-11-18 01:08:33.0 & EDA2                        & 42759 & ZHUHAI-1 02    & 161 & 162 & 633  & 1292 \\
146.1                        & 2021-11-18 01:08:33.0 & EDA2                        & 44881 & OBJECT C       & 136 & 111 & 948  & 1499 \\
146.1                        & 2021-11-18 01:08:33.0 & EDA2                        & 49399 & Z-SAT          & 117 & 133 & 797  & 1361 \\
146.1                        & 2021-11-18 01:08:33.0 & EDA2                        & 39444 & FUNCUBE 1      & 110 & 283 & 739  & 1539 \\
146.1                        & 2021-11-18 01:08:33.0 & EDA2                        & 43678 & DIWATA 2B      & 81  & 81  & 808  & 1422 \\
146.1                        & 2021-11-18 01:08:33.0 & EDA2                        & 43803 & JY1SAT         & 67  & 211  & 639 & 1411 \\
146.1                        & 2021-11-18 01:08:33.0 & EDA2                        & 47930 & HIROGARI       & 56  & 56  & 831  & 1026 \\
146.1                        & 2021-11-18 01:08:33.0 & EDA2                        & 40025 & QB50P1         & 38  & 38  & 803  & 1467 \\
\hline
\end{tabular}
\end{table*}
\clearpage
\begin{table*}[htbp]
\ContinuedFloat
\caption{Individual identifications per NORAD I.D. for each dataset (continued)}
\begin{tabular}{ccccccccc}

\hline
\hline
Frequency &
  \begin{tabular}[c]{@{}c@{}}Dataset\\ start date\end{tabular} &
  Telescope &
  \begin{tabular}[c]{@{}c@{}}NORAD\\ I.D.\end{tabular} &
  Name &
  \begin{tabular}[c]{@{}c@{}}Identifications\\ XX\end{tabular} &
  \begin{tabular}[c]{@{}c@{}}Identifications\\ YY\end{tabular} &
  \begin{tabular}[c]{@{}c@{}}Closest\\ range (km)\end{tabular} &
  \begin{tabular}[c]{@{}c@{}}Furthest\\ range (km)\end{tabular} \\
\hline
\hline

146.1                        & 2021-11-18 01:08:33.0 & EDA2                        & 44398 & OBJECT N       & 35  & 35  & 568  & 1529 \\
146.1                        & 2021-11-18 01:08:33.0 & EDA2                        & 44354 & PSAT 2         & 35  & 35  & 527  & 1287 \\
146.1                        & 2021-11-18 01:08:33.0 & EDA2                        & 43786 & ITASAT         & 34  & 39  & 719  & 1367 \\
146.1                        & 2021-11-18 01:08:33.0 & EDA2                        & 43770 & AO-95          & 22  & 24  & 606  & 1207 \\
146.1                        & 2021-11-18 01:08:33.0 & EDA2                        & 43780 & MOVE-II        & 21  & 17  & 651  & 1044 \\
146.1                        & 2021-11-18 01:08:33.0 & EDA2                        & 47950 & OBJECT U       & 19  & 18  & 603  & 1307 \\
146.1                        & 2021-11-18 01:08:33.0 & EDA2                        & 39427 & TRITON 1       & 12  & 12  & 1130 & 1275 \\
146.1                        & 2021-11-18 01:08:33.0 & EDA2                        & 44355 & BRICSAT 2      & 10  & 10  & 316  & 731  \\
146.1                        & 2021-11-18 01:08:33.0 & EDA2                        & 28650 & HAMSAT         & 0   & 32  & 826  & 901  \\
146.1                        & 2021-11-18 01:08:33.0 & EDA2                        & 47309 & CAPE-3         & 0   & 12  & 1028 & 1272 \\
\hline
159.4                        & 2020-06-26 11:48:10.0 & EDA2                        & 41999 & BGUSAT         & 469 & 515 & 491  & 1189 \\
159.4                        & 2020-06-26 11:48:10.0 & EDA2                        & 39427 & TRITON 1       & 240 & 237 & 662  & 1649 \\
159.4                        & 2020-06-26 11:48:10.0 & EDA2                        & 44426 & SWIATOWID      & 25  & 0   & 403  & 459  \\
159.4                        & 2020-06-26 11:48:10.0 & EDA2                        & 42778 & MAX VALIER SAT & 0   & 6   & 548  & 620  \\
\hline
159.4                        & 2020-06-26 12:34:07.0 & AAVS2                       & 41999 & BGUSAT         & 316 & 486 & 491  & 1183 \\
159.4                        & 2020-06-26 12:34:07.0 & AAVS2                       & 39427 & TRITON 1       & 158 & 180 & 659  & 1684 \\
\hline
159.4                        & 2021-11-16 01:24:18.0 & EDA2                        & 41999 & BGUSAT         & 70  & 77  & 701  & 1146 \\
159.4                        & 2021-11-16 01:24:18.0 & EDA2                        & 39427 & TRITON 1       & 6   & 6   & 608  & 612  \\
\hline
160.2                        & 2023-06-23 05:40:15.0 & EDA2                        & 56301 & STARLINK-30101 & 123 & 96  & 533  & 851  \\
160.2                        & 2023-06-23 05:40:15.0 & EDA2                        & 53807 & BLUEWALKER 3   & 112 & 88  & 526  & 1056 \\
160.2                        & 2023-06-23 05:40:15.0 & EDA2                        & 56292 & STARLINK-30098 & 68  & 62  & 531  & 689  \\
160.2                        & 2023-06-23 05:40:15.0 & EDA2                        & 56303 & STARLINK-30097 & 60  & 67  & 553  & 756  \\
160.2                        & 2023-06-23 05:40:15.0 & EDA2                        & 56293 & STARLINK-30090 & 58  & 63  & 529  & 697  \\
160.2                        & 2023-06-23 05:40:15.0 & EDA2                        & 56291 & STARLINK-30113 & 52  & 58  & 535  & 681  \\
160.2                        & 2023-06-23 05:40:15.0 & EDA2                        & 56289 & STARLINK-30095 & 49  & 66  & 475  & 745  \\
160.2                        & 2023-06-23 05:40:15.0 & EDA2                        & 56299 & STARLINK-30108 & 48  & 40  & 530  & 675  \\
160.2                        & 2023-06-23 05:40:15.0 & EDA2                        & 43834 & OBJECT D       & 42  & 110 & 766  & 1136 \\
160.2                        & 2023-06-23 05:40:15.0 & EDA2                        & 56300 & STARLINK-30104 & 32  & 22  & 529  & 605  \\
160.2                        & 2023-06-23 05:40:15.0 & EDA2                        & 56290 & STARLINK-30103 & 30  & 12  & 532  & 624  \\
160.2                        & 2023-06-23 05:40:15.0 & EDA2                        & 56302 & STARLINK-30111 & 29  & 33  & 562  & 621  \\
160.2                        & 2023-06-23 05:40:15.0 & EDA2                        & 56304 & STARLINK-30099 & 29  & 0   & 561  & 602  \\
160.2                        & 2023-06-23 05:40:15.0 & EDA2                        & 56839 & STARLINK-30076 & 27  & 0   & 584  & 723  \\
160.2                        & 2023-06-23 05:40:15.0 & EDA2                        & 56294 & STARLINK-30049 & 26  & 40  & 572  & 673  \\
160.2                        & 2023-06-23 05:40:15.0 & EDA2                        & 55712 & STARLINK-30037 & 24  & 12  & 368  & 542  \\
160.2                        & 2023-06-23 05:40:15.0 & EDA2                        & 56832 & STARLINK-30128 & 24  & 8   & 333  & 417  \\
160.2                        & 2023-06-23 05:40:15.0 & EDA2                        & 56306 & STARLINK-30112 & 24  & 0   & 561  & 606  \\
160.2                        & 2023-06-23 05:40:15.0 & EDA2                        & 55695 & STARLINK-30050 & 20  & 48  & 363  & 639  \\
160.2                        & 2023-06-23 05:40:15.0 & EDA2                        & 56826 & STARLINK-30114 & 19  & 29  & 417  & 519  \\
160.2                        & 2023-06-23 05:40:15.0 & EDA2                        & 56296 & STARLINK-30089 & 16  & 26  & 449  & 574  \\
160.2                        & 2023-06-23 05:40:15.0 & EDA2                        & 56838 & STARLINK-30123 & 15  & 23  & 458  & 722  \\
160.2                        & 2023-06-23 05:40:15.0 & EDA2                        & 56833 & STARLINK-30131 & 12  & 18  & 357  & 471  \\
160.2                        & 2023-06-23 05:40:15.0 & EDA2                        & 56699 & STARLINK-30127 & 12  & 17  & 534  & 590  \\
160.2                        & 2023-06-23 05:40:15.0 & EDA2                        & 56700 & STARLINK-30053 & 12  & 0   & 529  & 570  \\
160.2                        & 2023-06-23 05:40:15.0 & EDA2                        & 56844 & STARLINK-30136 & 11   & 15 & 498  & 684  \\
160.2                        & 2023-06-23 05:40:15.0 & EDA2                        & 56298 & STARLINK-30109 & 0   & 17  & 615  & 795  \\
160.2                        & 2023-06-23 05:40:15.0 & EDA2                        & 56693 & STARLINK-30067 & 0   & 14  & 535  & 557  \\
160.2                        & 2023-06-23 05:40:15.0 & EDA2                        & 39427 & TRITON 1       & 0   & 6   & 1533 & 1573 \\
\hline
185.2                        & 2020-02-07 10:44:49.3 & EDA2                    & 41999 & BGUSAT         & 90  & 98  & 494  & 1180 \\
\hline

\end{tabular}
\end{table*}
\clearpage
\begin{table*}[!t]
\ContinuedFloat
\caption{Individual identifications per NORAD I.D. for each dataset (continued)}
\begin{tabular}{ccccccccc}

\hline
\hline
Frequency &
  \begin{tabular}[c]{@{}c@{}}Dataset\\ start date\end{tabular} &
  Telescope &
  \begin{tabular}[c]{@{}c@{}}NORAD\\ I.D.\end{tabular} &
  Name &
  \begin{tabular}[c]{@{}c@{}}Identifications\\ XX\end{tabular} &
  \begin{tabular}[c]{@{}c@{}}Identifications\\ YY\end{tabular} &
  \begin{tabular}[c]{@{}c@{}}Closest\\ range (km)\end{tabular} &
  \begin{tabular}[c]{@{}c@{}}Furthest\\ range (km)\end{tabular} \\
\hline
\hline

229.7                        & 2021-11-16 01:10:45.0 & AAVS2                        & 37153 & STRELA 3       & 36  & 32  & 1541 & 2975 \\
229.7                        & 2021-11-16 01:10:45.0 & AAVS2                        & 28420 & COSMOS 2409    & 35  & 27  & 1806 & 2886 \\
229.7                        & 2021-11-16 01:10:45.0 & AAVS2                        & 35500 & COSMOS 2453    & 35  & 17  & 1598 & 3013 \\
229.7                        & 2021-11-16 01:10:45.0 & AAVS2                        & 27868 & COSMOS 2400    & 32  & 36  & 1882 & 2969 \\
229.7                        & 2021-11-16 01:10:45.0 & AAVS2                        & 32954 & COSMOS 2437    & 27  & 11  & 1553 & 2952 \\
229.7                        & 2021-11-16 01:10:45.0 & AAVS2                        & 27059 & GONETS D1 8    & 26  & 5   & 1535 & 2843 \\
229.7                        & 2021-11-16 01:10:45.0 & AAVS2                        & 38733 & COSMOS 2481    & 24  & 21  & 1727 & 2935 \\
229.7                        & 2021-11-16 01:10:45.0 & AAVS2                        & 27056 & COSMOS 2385    & 23  & 23  & 1502 & 2879 \\
229.7                        & 2021-11-16 01:10:45.0 & AAVS2                        & 27465 & COSMOS 2391    & 16  & 26  & 1520 & 2911 \\
229.7                        & 2021-11-16 01:10:45.0 & AAVS2                        & 32956 & COSMOS 2439    & 12  & 0   & 2081 & 2871 \\
229.7                        & 2021-11-16 01:10:45.0 & AAVS2                        & 40554 & GONETS M 13    & 9   & 0   & 1510 & 2127 \\
229.7                        & 2021-11-16 01:10:45.0 & AAVS2                        & 44906 & GONETS M 15    & 7   & 6   & 1642 & 2132 \\
229.7                        & 2021-11-16 01:10:45.0 & AAVS2                        & 47229 & GONETS M 22    & 6   & 5   & 1713 & 2125 \\
229.7                        & 2021-11-16 01:10:45.0 & AAVS2                        & 39249 & GONETS M-5     & 6   & 0   & 1536 & 2055 \\
229.7                        & 2021-11-16 01:10:45.0 & AAVS2                        & 46486 & GONETS M 17    & 5   & 0   & 1602 & 2040 \\
\hline

\end{tabular}
\end{table*}

\end{document}